\documentclass[11pt]{article}
\pdfoutput=1
\usepackage{jheppub}
\usepackage{amsmath}
\usepackage{cases}
\usepackage{amssymb}
\usepackage[retainorgcmds]{IEEEtrantools}
\usepackage{hyperref}
\usepackage{graphicx}
\usepackage[justification=justified,font=small,labelfont=bf]{caption}
\usepackage{subcaption}
\usepackage[shortlabels]{enumitem}
\usepackage[utf8]{inputenc}

\newcommand{\ie}{\textit{i.e.}\ }
\newcommand{\eg}{\textit{e.g.}\ }
\newcommand{\viz}{\textit{viz.}\ }

\title{Barnacles and Gravity}
\date{} 

\begin{document}

\author{James H. C. Scargill}

\affiliation{Center for Quantum Mathematics and Physics (QMAP), Department of Physics, University of California at Davis, One Shields Avenue, Davis, CA 95616, USA} 

\emailAdd{jhcscargill@gmail.com}

\abstract{
Theories with more than one vacuum allow quantum transitions between them, which may proceed via bubble nucleation; theories with more than two vacua posses additional decay modes in which the wall of a bubble may further decay. The instantons which mediate such a process have $O(3)$ symmetry (in four dimensions, rather than the usual $O(4)$ symmetry of homogeneous vacuum decay), and have been called `barnacles'; previously they have been studied in flat space, in the thin wall limit, and this paper extends the analysis to include gravity.
It is found that there are regions of parameter space in which, given an initial bubble, barnacles are the favoured subsequent decay process, and that the inclusion of gravity can enlarge this region.
The relation to other heterogeneous vacuum decay scenarios, as well as some of the phenomenological implications of barnacles are briefly discussed.
}

\setcounter{tocdepth}{2}

\maketitle



\section{Introduction}

The topic of vacuum decay in quantum field theory has a long history, dating back to the pioneering work of Coleman and collaborators \cite{Coleman:1977py, Callan:1977pt, Coleman:1980aw}. The essential fact is that in a theory with more than one vacuum (\ie the potential energy has more than one local minimum), there will be transitions between them, with a rate $\Gamma \sim \mathrm{e}^{-B}$, where $B$ is the difference in Euclidean action between the final and initial field configurations; these configurations extremise the Euclidean action and hence are instantons. In cases where the initial state is a homogeneous region of one vacuum, the final state typically consists of a bubble of the new vacuum in a `sea' of the old one,\footnote{In the presence of gravity and positive vacuum energies there exists another type of decay, the Hawking-Moss instanton \cite{Hawking:1981fz}, which can be interpreted as the simultaneous transition of a whole horizon volume to the new vacuum \cite{Brown:2007sd} rather than nucelation of a bubble. In this paper I will assume such processes are subdominant.} and in the absence of gravity it has been proven that the difference in Euclidean action (\ie $B$) is minimised when this bubble has $O(4)$ symmetry (in four spacetime dimensions---in which I will work throughout this paper) \cite{Coleman:1977th, Blum:2016ipp}.\footnote{This changes in flat space in the presence of a non-zero temperature, in which case the minimum action configuration generally has less symmetry.} Although not proven, this is expected to hold true when gravity is included.

On the other hand, in everyday bubble nucleation (\eg Champagne) it is generally not the homogoeneous decay rate which is important but the heterogeneous one, since impurities act as seeds and enhance the rate. This has led to investigations of a similar effect in QFT bubble nucleation. For example \cite{Grinstein:2015jda} studied in flat spacetime the effect of impurities much larger than than the bubble size, whilst the possibility of black holes acting as nucleation sites has been studied by many authors, and most recently by \cite{Gregory:2013hja}.

An intriguing possibility then comes to mind: that in a theory with more than two vacua, a vacuum bubble itself may act as a seed for further decay. This was studied in flat space, in the thin wall limit,\footnote{In this limit, which is realised when the height of the barrier between two vacua is much larger than their energy separation, a bubble consists of regions in which the field is exactly in a vacuum, separated by a wall parametrised just by its tension; going beyond this, even in the $O(4)$ symmetric case, generally allows only numerical solutions.} in \cite{Balasubramanian:2010kg, Czech:2011aa}, and the resulting instantons were named `barnacles,' since one vacuum region grips onto another, like a barnacle on the hull of a cosmic ship traversing the multiverse.

The spectrum of fluctuations around the instanton describing a barnacle has two negative modes. On the face of it this is rather worrying, since Coleman showed long ago that only instantons with exactly one negative eigenvalue are relevant for the decay of empty space \cite{Coleman:1987rm}. However, as first explained in \cite{Czech:2011aa}---which discussion is recapitulated below---this turns out not to be a problem, and even is to be expected, since a barnacle instanton really describes \emph{two} decay processes: the nucleation of an initial, spherical bubble, followed by the decay of a section of its wall.

Another feature of barnacles is that they do not have $O(4)$ symmetry, since the barnacle itself\footnote{Note that I will use the word barnacle to describe both the entire instanton, and the individual vacuum region which results from the decay of the wall of a bubble, however the context should always make clear what is meant.} picks out a particular position on the wall, but there remains a residual $O(3)$ symmetry around the axis going through this point.

As well as being intrinsically interesting due to being `non-standard' tunnelling events, barnacles may be relevant in the context of creating new universes via tunnelling in a landscape (be it from string theory or otherwise) since such scenarios have more than two vacua. Also, one would also generically expect the effects of gravity to be important in such a case, motivating their inclusion here.

The structure of the rest of this paper is as follows. The next subsection explains in a little more detail how instantons with multiple negative modes can be understood to mediate multiple decay events; section \ref{sec-flat space barnacles} covers the results of \cite{Balasubramanian:2010kg} concerning barnacles in flat space, after which the main result of this paper---the calculation of the action for a barnacle instanton in the presence of gravity---is described in section \ref{sec-barnacles and gravity}; in section \ref{sec-comp decay rates} the behaviour of the action (and hence approximately the rate) for barnacles is examined and it is compared with the action for configurations with two spherical bubbles; section \ref{sec-obs conqs} briefly discusses some of the observational and phenomenological consequences of barnacles; finally section \ref{sec-conc} offers some conclusions; an appendix contains some additional details for the calculation of the action.

\subsection{Instantons with Multiple Negative Modes} \label{sec-multi neg}

As mentioned above, the second variation of the action describing a barnacle has two negative eigenvalues. As shown in \cite{Czech:2011aa}, and briefly sketched below, instantons with multiple negative eigenvalues can be understood to describe multiple decay processes.

Consider a theory with three vacua, $A$, $B$, and $C$, and evaluate the Euclidean partition function in the saddle point approximation around the $A$ vacuum:
\begin{equation}
Z = \mathrm{e}^{-S_A} = \mathrm{e}^{-V_A \mathrm{Vol}_A},
\end{equation}
where $V_A$ is the energy density of the $A$ vacuum, the volume of which region is $\mathrm{Vol}_A$. The action for an $AB$ instanton, again evaluated in the saddle point approximation, is
\begin{equation}
I_{AB} = c_{AB} \left[ {\det}' S_{AB}'' \right]^{-\frac{1}{2}} \mathrm{e}^{-S_{AB}} \mathrm{Vol}_A,
\end{equation}
where $c_{AB}$ collects factors unimportant for this analysis, and the prime denotes that the determinant excludes the zero eigenvalues. In the dilute gas approximation these instantons correct the partition function in the following way:
\begin{equation}
Z \to \mathrm{e}^{-V_A \mathrm{Vol}_A} \sum_{n=0}^\infty \frac{\left( I_{AB} \right)^n}{n!} = \mathrm{e}^{-(V_A + i \Gamma_{AB}) \mathrm{Vol}_A},
\end{equation}
where
\begin{equation}
\Gamma_{AB} = c_{AB} \left[ - {\det}' S_{AB}'' \right]^{-\frac{1}{2}} \mathrm{e}^{-S_{AB}}.
\end{equation}
Given that there is one negative mode of fluctuations about $S_{AB}$, one sees that $V_A$ has acquired an imaginary part, which can be interpreted as a decay rate.

Now include $BC$ instantons:
\begin{equation}
Z \to \mathrm{e}^{-V_A \mathrm{Vol}_A} \sum_{n=0}^\infty \frac{\left( i \Gamma_{AB} \sum_m \frac{\left( I_{BC} \right)^m}{m!} \mathrm{Vol}_A\right)^n}{n!},
\end{equation}
and note that in thin-wall limit 
\begin{equation}
S_{AB} = - (V_A - V_B) \mathrm{Vol}_B + \sigma_{AB} \mathrm{Vol}_{AB},
\end{equation}
where $\mathrm{Vol}_{AB}$ is the volume of the wall separating the $A$ and $B$ vacuum regions, which has tension $\sigma_{AB}$. Thus one has
\begin{equation}
\Gamma_{AB} \to c_{AB} \left[ - {\det}' S_{AB}'' \right]^{-\frac{1}{2}} \mathrm{e}^{(V_A - \left[ V_B + i \Gamma_{BC} \right]) \mathrm{Vol}_B - \sigma_{AB} \mathrm{Vol}_{AB} },
\end{equation}
where
\begin{equation}
\Gamma_{BC} = c_{BC} \left[ - {\det}' S_{BC}'' \right]^{-\frac{1}{2}} \mathrm{e}^{-S_{BC}};
\end{equation}
\ie the $B$ vacuum has now acquired a decay rate.

Finally, include barnacles:
\begin{align}
&Z \to \mathrm{e}^{-V_A \mathrm{Vol}_A} \sum_{n=0}^\infty \frac{\left( i \Gamma_{AB} \sum_m \frac{\left( c_b \left[ \det' \tilde{S}_b'' \right]^{-\frac{1}{2}} \mathrm{e}^{-\tilde{S}_b}  \mathrm{Vol}_{AB}\right)^m}{m!} \mathrm{Vol}_A\right)^n}{n!}, \\
&\Gamma_{AB} \to c_{AB} \left[ - {\det}' S_{AB}'' \right]^{-\frac{1}{2}} \mathrm{e}^{(V_A - V_B) \mathrm{Vol}_B - \left[\sigma_{AB} + i \Gamma_b \right] \mathrm{Vol}_{AB} },
\end{align}
where
\begin{equation}
\Gamma_{b} = c_b \left[ - {\det}' \tilde{S}_b'' \right]^{-\frac{1}{2}} \mathrm{e}^{-\tilde{S}_b},
\end{equation}
and $\tilde{S}_b \equiv S_b - S_{AB}$ is the difference in Euclidean action between an $AB$ bubble dressed with a barnacle and the $AB$ bubble alone. In this final case it is the $AB$ wall tension which has acquired an imaginary part, and hence the barnacle instanton can be interpreted as mediating decay of the $AB$ wall.

\section{Barnacles in Flat Space} \label{sec-flat space barnacles}

\begin{figure}
\centering
\includegraphics[width=0.6\textwidth]{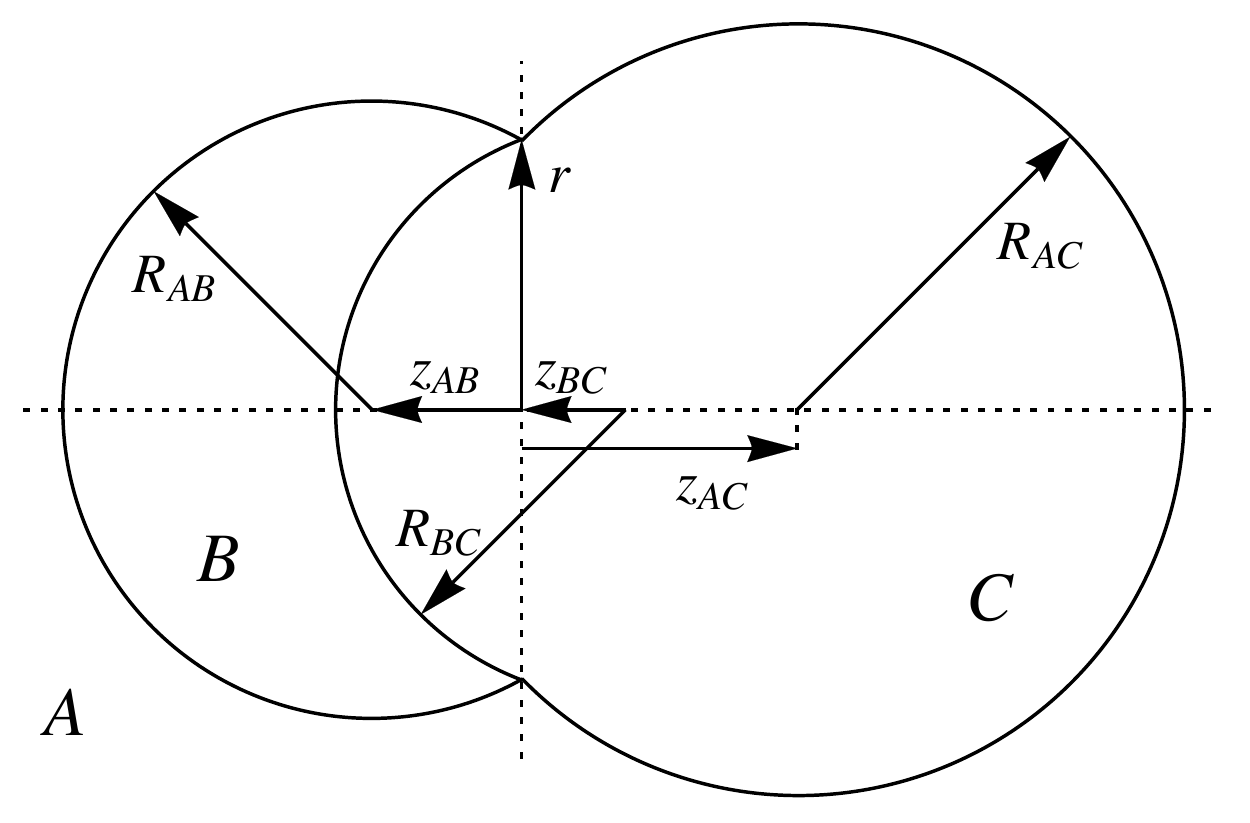}
\caption{The flat-space barnacle geometry; $z_{AC}$ is measured increasing to the right, whilst $z_{AB}$ and $z_{BC}$ are measured increasing to the left, with the zero point at the plane which contains the junction point where all three vacua meet. Two angular directions have been suppressed, so each point is actually a two-sphere.}\label{fig-barnacle_flat_space}
\end{figure}

This section briefly recapitulates the results, from \cite{Balasubramanian:2010kg}, concerning the action for a barnacle in flat space. The geometry is shown in figure \ref{fig-barnacle_flat_space}, and the action is
\begin{equation}
S_b = -\sum_{i \in \{A,B,C\}} (V_A - V_i) \mathrm{Vol}_i + \sum_{X \in \{AB,AC,BC\}} \sigma_X \mathrm{Vol}_X + \mu \mathrm{Vol}_J,
\end{equation}
where $i$ labels the vacuum regions, $X$ the walls, and $J$ denotes the junction (two-sphere) at which the three vacua meet; $\mu$ is a parameter that depends on how the field interpolates between the three vacua at the junction (much like how the wall tension depends on how the field interpolates between two vacua), and it calculation is discussed later in this section. Extremising the action one finds
\begin{equation}
S_b =S_{AB} \, k\left( \frac{z_{AB}}{R_{AB}} \right) + S_{AC} \, k\left( \frac{z_{AC}}{R_{AC}} \right) + S_{BC} \, k\left( \frac{z_{BC}}{R_{BC}} \right) + 4 \pi r^2 \mu, \label{Sbarn flat space}
\end{equation}
where
\begin{equation}
k(x) = \frac{1}{2} + \frac{1}{\pi}x(2x^2-1) \sqrt{1-x^2} + \frac{1}{\pi} \sin^{-1} x,
\end{equation}
varies between $0$ and $1$, and $S_X$ is the thin wall action for an $O(4)$ symmetric bubble:
\begin{equation}
S_X = \frac{27\pi^2}{2} \frac{\sigma_X^4}{\epsilon_X^3},
\end{equation}
and $\epsilon_{ft} = V_f - V_t$. The bubble segment radii take their usual thin wall values:
\begin{equation}
R_X = \frac{3 \sigma_X}{\epsilon_X},
\end{equation}
and the magnitudes of the $z$'s are constrained to satisfy $z_X^2 + r^2 = R_X^2$. Their signs, and the value of $r$ can be determined from
\begin{equation}
r \sum_X \epsilon_X z_X = 6 \mu. \label{flat space r eqn}
\end{equation}
For a given set of vacuum energies and wall tensions, the left hand side of this equation is clearly bounded, and hence there exists a maximum value of $|\mu|$ beyond which it is no longer possible to find a barnacle. For $\mu = 0$, however, a barnacle always exists, and one has
\begin{equation}
r^2 = \left(\frac{3}{2}\right)^2 \frac{2\sigma_{AB}^2\sigma_{BC}^2 + 2\sigma_{BC}^2\sigma_{AC}^2 + 2\sigma_{AC}^2\sigma_{BC}^2 - \sigma_{AB}^4 - \sigma_{BC}^4 - \sigma_{AC}^4}{\epsilon_{AB}^2\sigma_{BC}^2 + \epsilon_{BC}^2\sigma_{AB}^2 + \epsilon_{AB}\epsilon_{BC}(\sigma_{AB}^2 + \sigma_{BC}^2 - \sigma_{AC}^2)}. \label{r2 flat space}
\end{equation}

Looking at the geometry depicted in figure \ref{fig-barnacle_flat_space}, one may worry that the above calculation of the barnacle decay rate is only valid in the limit that the barnacle appears at the same time as the initial seed bubble. On the other hand, the Lorentz invariance of a spherical bubble wall is the reason why one does not integrate over all possible boosts and rotations of the bubble when calculating the decay rate (na\"ively yielding an infinite rate), and given that the barnacle is no longer $SO(1,3)$ invariant one may wonder what happens to this na\"ive infinity. 

These two issues turn out to have the same resolution. The initial bubble wall is given by the hyperboloid $\mathbf{x}^2 - t^2 = R_X^2$, which is $SO(1,3)$ invariant, and so a geometry with the barnacle produced at the same time as the initial bubble can be transformed into one with the barnacle at any other location (in space and time) on the bubble wall simply by a Lorentz transformation---yielding the same probability. Similarly, integrating over boosts and rotations of the initial bubble would really be just the same as integrating over all possible positions on the (infinite) bubble wall worldvolume for the barnacle to appear.

\subsection{Determining the Wall Tensions and Energy Density at the Junction}

Given a set of fields $\phi^i$, and a (constant) field space metric $k_{ij}$ their equation of motion is\footnote{The wall tension and junction energy density should only depend on the short distance features of the solution and thus this section also applies to the case in which gravity is included.}
\begin{equation}
k_{ij} \partial_\mu \partial^\mu \phi^j = \frac{\partial V}{\partial \phi^i}.
\end{equation}
Integrating this with respect to $\phi^i$ yields
\begin{equation}
\frac{1}{2} k_{ij} \partial_\mu \phi^i \partial^\mu \phi^j = V(\phi) - V_\text{fv}.
\end{equation}

The energy of the wall per unit area (\ie its tension) is thus given by
\begin{align}
\sigma &= \int \mathrm{d}s\, \left( \frac{1}{2} k_{ij} \partial_\mu \phi^i \partial^\mu \phi^j + V - V_\text{fv} \right) \nonumber \\
& = \int \mathrm{d}l\, \sqrt{2 \left(V - V_\text{fv}\right)}, \label{sigma equation}
\end{align}
where $s$ is the coordinate transverse to the wall and $\mathrm{d}l^2 = k_{ij} \mathrm{d}\phi^i \mathrm{d}\phi^j$ is the line element in field space; the integration is taken over the path in field space with minimises the integral, from the initial false vacuum point, to the classical escape point (which in the thin wall limit is taken to be the true vacuum point).

Expression \eqref{sigma equation} means that any three wall tensions must satisfy a triangle inequality:
\begin{equation}
\sigma_{ij} < \sigma_{ik} + \sigma_{kj}, \label{triangle inequality}
\end{equation}
since the path from $i$ to $j$ going via $k$ would saturate this.

A compelling aspect of \eqref{sigma equation} is that one can calculate the wall tension without requiring a bounce solution.
The same is not true however for the energy density at the junction. The junction has the geometry of a two-sphere, and integrating transverse to this one has
\begin{align}
\mu &= \int \mathrm{d}x_1\mathrm{d}x_2\, \left( \frac{1}{2} k_{ij} \partial_\mu \phi^i \partial^\mu \phi^j + V - V_\text{fv} \right) - \sum_X \int \mathrm{d}x^\parallel_X\, \sigma_X \nonumber \\
&= \int \mathrm{d}x_1\mathrm{d}x_2\, 2 \left(V - V_\text{fv}\right) - \sum_X \int \mathrm{d}x^\parallel_X\, \sigma_X \label{mu x equation} \\
&= \int \mathrm{d}\phi_1 \mathrm{d}\phi_2 \frac{2 \left(V - V_\text{fv}\right)}{\left| \partial_\mu \phi_i \right|} - \sum_X \int \mathrm{d}\phi^\parallel_X\, \frac{\sigma_X}{\partial_{x_X^\parallel} \phi_X^\parallel}, \label{mu phi equation}
\end{align}
where $x_X^\parallel$ is the coordinate parallel to the $X$ wall, and the sum is over the walls which are incident at the junction; in the last line I have specialised to the case of two fields. Note that to properly determine $\mu$ one must subtract off the energy of the walls, which is already included in the action through the terms $(\text{Area of walls}) \times \sigma$, just as to properly determine $\sigma$ one has to subtract off the energy of the vacuum regions (through $V \to V-V_\text{fv}$), which is already included in the action through the terms $(\text{Volume of vacuum regions}) \times V$.

Unlike the field space path, determining the Jacobian $\left| \partial_\mu \phi_i \right|$ (along with the quantities $\partial_{x_X^\parallel} \phi_X^\parallel$) does not seem to be possible without possessing a bounce solution. That being said, given the earlier comment that the junction energy density should only depend on the short distance features of the solution, one expects that a two dimensional solution, in which only the directions transverse to the junction two sphere are retained, should suffice.

It is worth commenting on the sign of $\mu$. The wall tensions must be positive, since the potential between two vacua must at some point rise above the false vacuum, and hence even when subtracting off the false vacuum energy, the result (which one integrates between the two vacua) will still be positive. Similarly the potential in between three vacua must be larger than the potential on the tunnelling paths, and so one may expect that $\mu$ must be positive. On the other hand, when subtracting off the wall tension there will be regions involving more than one subtraction, which makes it seem possible that $\mu$ could be negative.

As an explicit example, consider the following potential with four, degenerate vacua:
\begin{equation}
V(\phi_1, \phi_2) = \lambda_1 \left(\phi_1^2 - v_1^2\right)^2 + \lambda_2 \left( \phi_2 - v_2^2 \right)^2;
\end{equation}
clearly at any point in between the vacua, \ie $\{\phi_1, \phi_2\} \in \{(-v_1, v_1),(-v_2,v_2)\}$, its value is greater than on a tunnelling path, \ie $|\phi_i| = v_i$, yet, due to its separable structure, it is easy to determine that for a solution in which the field is in a different vacuum in each quadrant of the $(x,y)$ plane, that the parameter $\mu$ actually vanishes.

Therefore, in the absence of conclusive evidence that $\mu$ cannot be negative, I will allow it take either sign.

\section{Barnacles and Gravity} \label{sec-barnacles and gravity}

\begin{figure}[tp]
\begin{subfigure}{0.4\textwidth}
\includegraphics[width=\textwidth]{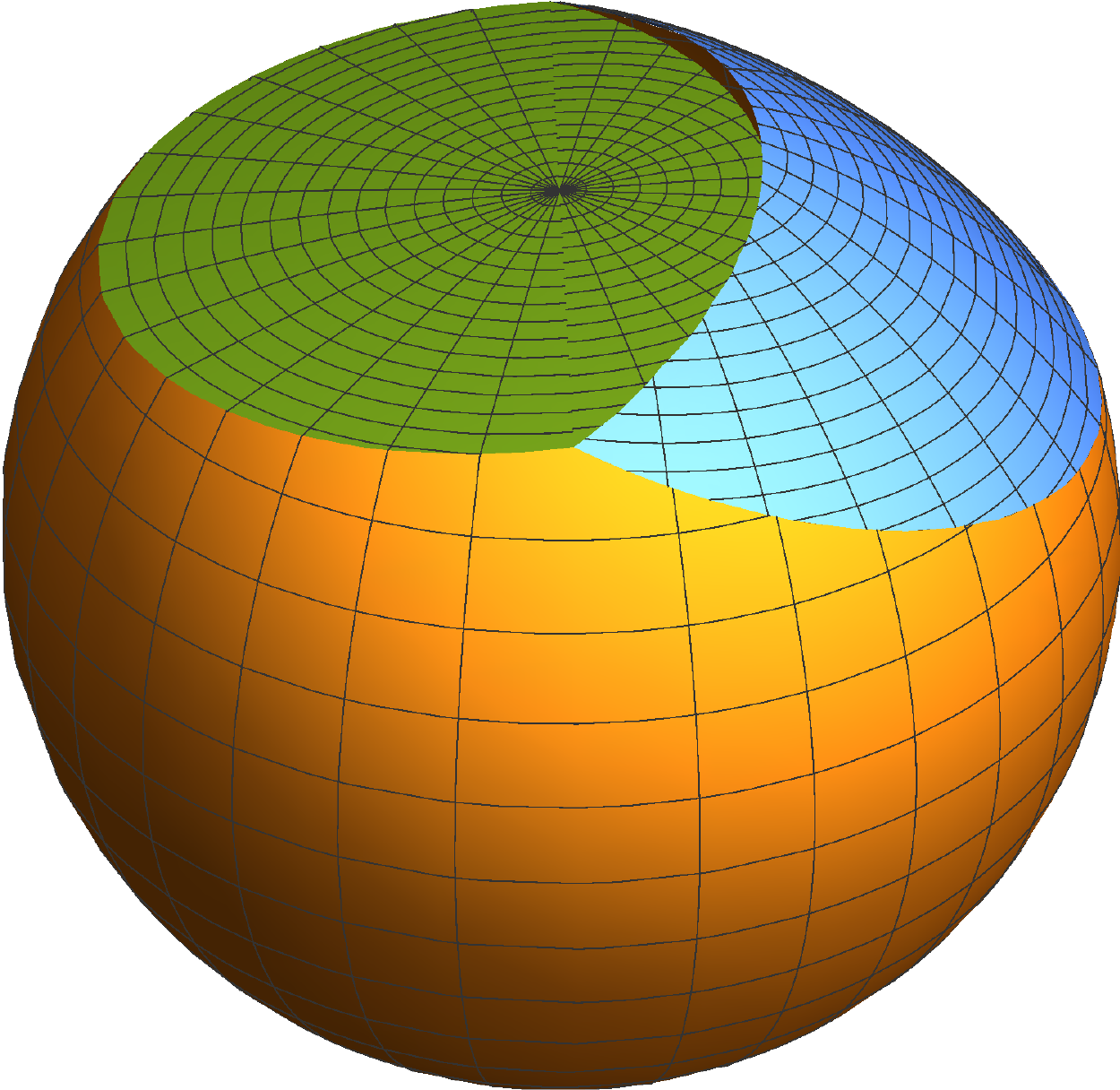}
\end{subfigure}
\hfill
\begin{subfigure}{0.4\textwidth}
\includegraphics[width=\textwidth]{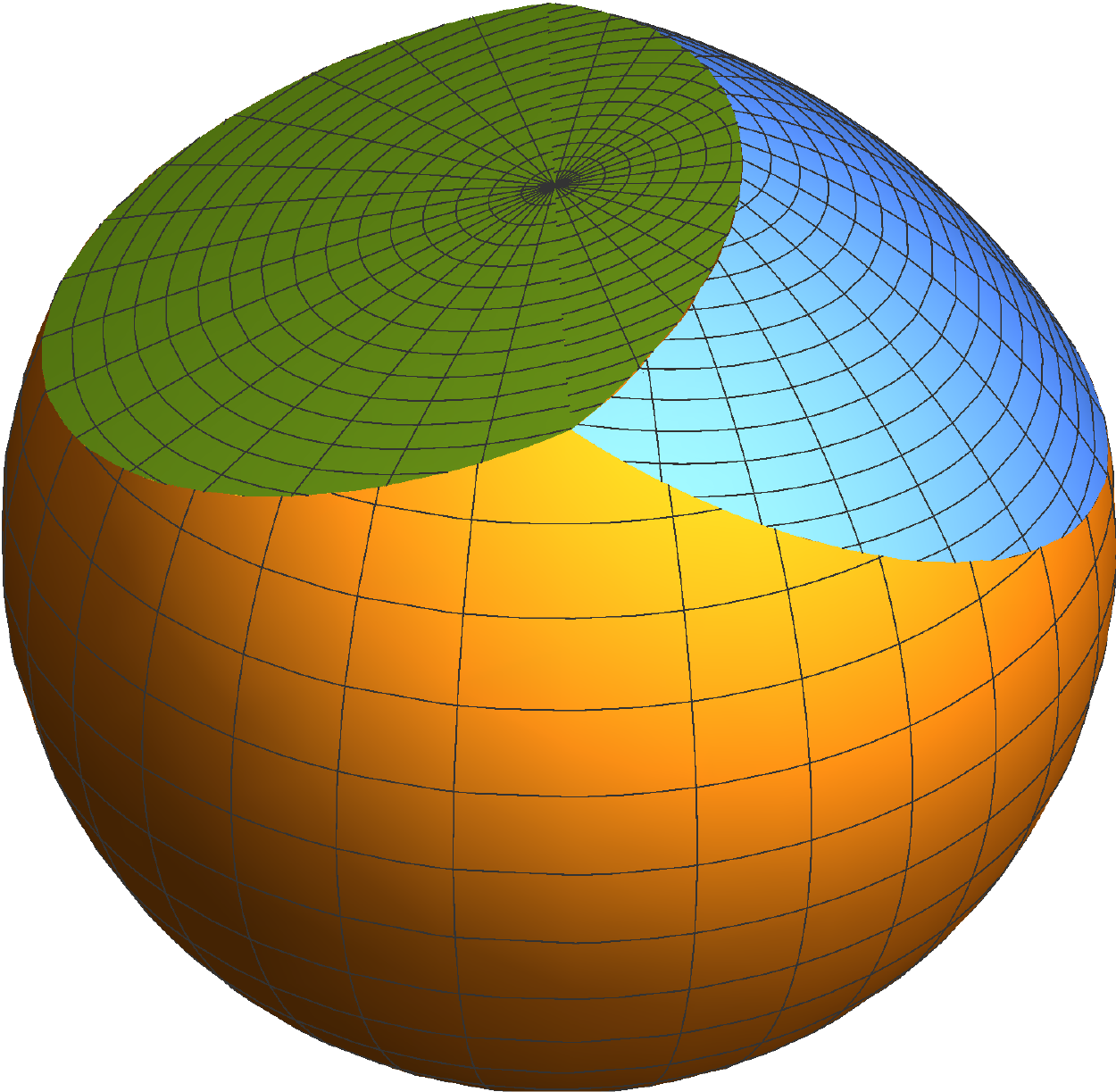}
\end{subfigure}
\caption{Examples of barnacles when gravity is included, the parameters are: $V_A = 1$, $V_B = 0.5$, $V_C = 0.01$, $\sigma_{AB} = 0.3$, $\sigma_{AC} = 0.5$, $\sigma_{BC} = 0.3$, in units in which $\kappa = 1$; two of the angular directions have been suppressed, so each point is a two-sphere. Orange represents the false vacuum, blue is intermediate vacuum, and green is true vacuum. The one on the left has $\mu = 0$ and hence no deficit angle at the junction point where the three vacua meet, whilst the one on the right has $\mu = 0.1$; note that the deficit angle is not equal to the angle through which the embedding in $\mathbb{R}^3$ fails to be continuous, which one sees is non-zero in the left hand image.}\label{fig-barnacle examples}
\end{figure}

Let us now come to the task of including the effects of gravity on barnacles. Remaining within the thin wall regime, the vacuum regions become portions of de Sitter space\footnote{With a view to phenomenology, I take all vacuum energies to be positive, however the limit in which the true vacuum is Minkowski is easily taken.}---which in four Euclidean dimensions is a four-sphere---separated by domain walls, as depicted in figure \ref{fig-barnacle examples}. For concreteness, I write the metric in Euclidean de Sitter space in the presence of vacuum energy $V$ as
\begin{equation}
\mathrm{d}s^2 = \frac{3}{\kappa V} \left[ \mathrm{d}\xi^2 + \sin^2 \xi \left( \mathrm{d}\psi^2 + \sin^2\psi\, \mathrm{d}\Omega_2^2 \right) \right],
\end{equation}
where $\kappa = 8 \pi G = M_\text{Pl}^{-2}$ is the reduced Planck mass, $\mathrm{d}\Omega_2^2$ is the metric on a unit two-sphere, and the fields will only depend on the $\xi$ and $\psi$ coordinates.

The action takes the form
\begin{align}
S_b = &\sum_i \left[ \int_{\mathrm{Vol}_i} \mathrm{d}^4x\, \sqrt{|g|} \left( V_i - \frac{1}{2\kappa} \mathcal{R} \right) -\frac{1}{\kappa} \int_{\partial \mathrm{Vol}_i} \mathrm{d}^3y\, \sqrt{|\gamma|}\, \mathcal{K} \right] \nonumber\\ 
&+ \sum_{X} \int_{(\partial \mathrm{Vol})_X} \mathrm{d}^3y\, \sqrt{|\gamma|}\, \sigma_X + \int_J \mathrm{d}^2z \sqrt{|h|} \left( \mu - \frac{\Delta}{\kappa} \right) \label{Sbarn with grav} \\
&- \left(-\frac{24\pi^2}{\kappa^2 V_A}\right) \nonumber
\end{align}
where $i \in \{A,B,C\}$ runs over the vacua, $X \in \{AB,BC,AC\}$ runs over the walls, and $J$ is the junction two sphere where the three vacua meet. $\mathcal{R}$ is the Ricci scalar in each vacuum region, and $\mathcal{K}$ is the extrinsic curvature of its boundary, whilst $\gamma$ and $h$ denote the induced metrics on the boundaries and junction point respectively.
The penultimate term requires a little discussion: the energy density associated with $\mu$ induces a conical singularity at the junction point, and the contribution of the Ricci scalar at a conical singularity of deficit angle $\Delta$ is given by $2 \Delta \delta_J$, where the delta function satisfies $\int \mathrm{d}^4x\, \sqrt{|g|} \delta_J = \int_J \mathrm{d}^2z \sqrt{|h|}$ \cite{Fursaev:1995ef}.\footnote{One would also get the same form for the junction by considering the `corner' terms, associated with the intersection of the boundaries $\partial Vol_i$ (since the extrinsic curvatures are infinite there), which would take the form $\frac{-1}{\kappa}\int_J \mathrm{d}^2z \sqrt{|h|} (2\pi - \alpha_A - \alpha_B - \alpha_C)$ \cite{Hayward:1993my}, where $\alpha_i$ is the angle between the two portions of the boundary of the region of vacuum $i$, and by the definition of the deficit angle one has $\alpha_A + \alpha_B + \alpha_C = 2\pi - \Delta$.}
The final term is simply subtracting off the action of the initial homogeneous false vacuum region.

\subsection{Calculating the Action}

The intrinsic and extrinsic curvatures of each bubble segment and its boundary are coordinate scalars, and thus each can be calculated in a coordinate system which is centred on that particular bubble segment, yielding the same values as in the $O(4)$-symmetric case, \ie
\begin{equation}
\mathcal{R} = 4 \kappa V, \qquad \mathcal{K} = \sqrt{3 \kappa V} \cot R.
\end{equation}
In the $O(4)$-symmetric case, in the coordinate system on the false vacuum side, the bubble wall is simply at $\xi_f = R$, which is then matched on the other side by requiring the metric to be continuous: $\frac{3}{\kappa V_t} \sin^2 \xi_t = \frac{3}{\kappa V_f} \sin^2 R$.

\begin{figure}
\centering
\includegraphics[scale=0.5]{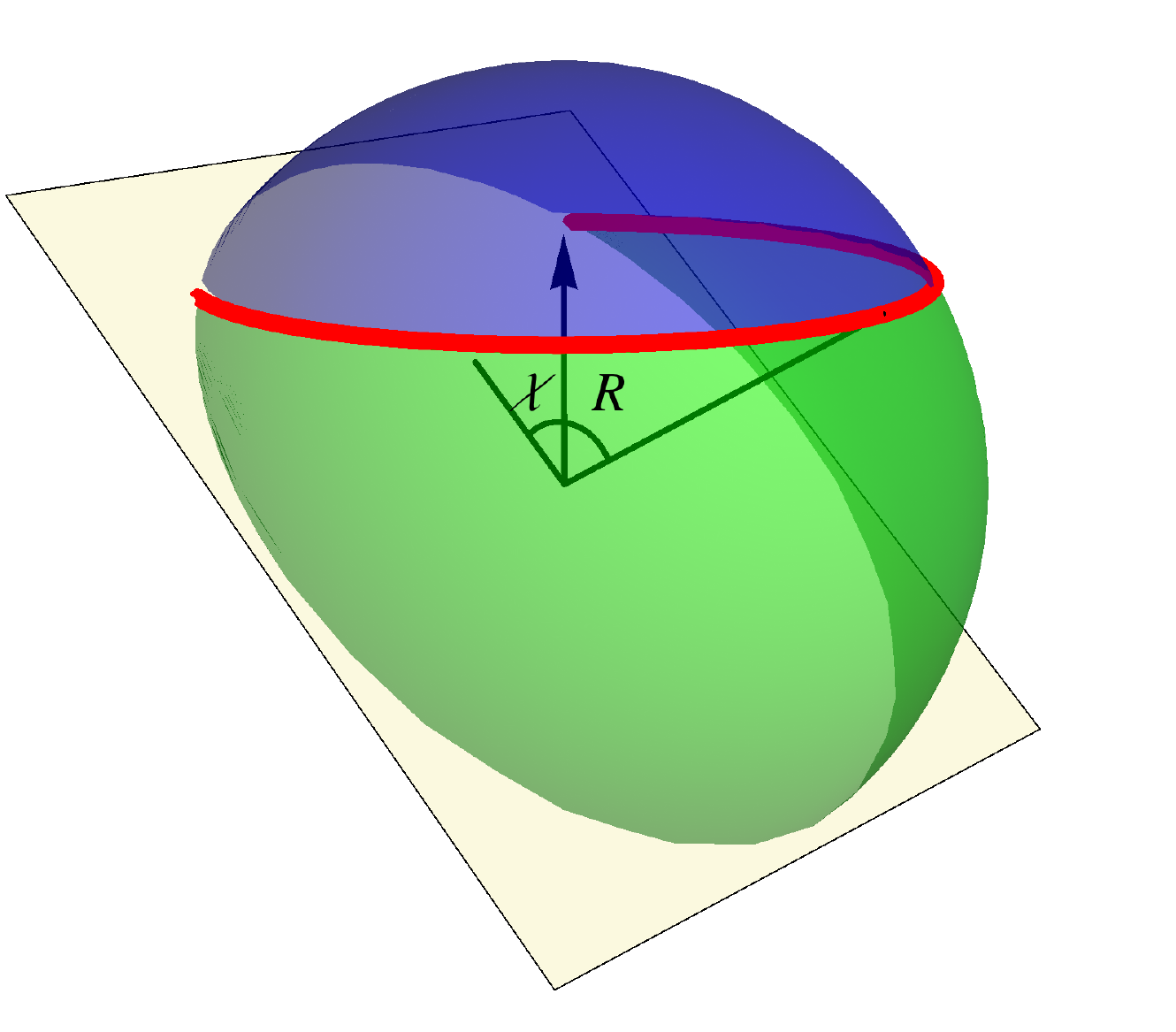}
\caption{The upper, (translucent) blue section is \eqref{bubble segment volume}; the red line is \eqref{area function}.} \label{fig-Vol_calculation}
\end{figure}

The volume of a segment of a bubble of coordinate radius $R$, and the area of its boundary, as depicted in blue and red respectively in figure \ref{fig-Vol_calculation}, can be calculated in the following way. Consider a sphere embedded in Euclidean space, bisected by a plane at a angle $\chi$ to the vertical; their intersection describes a circle $y^2 + z^2 \sec^2 \chi = 1$, $x = -z\tan\chi$, which upon passing to spherical coordinates becomes
\begin{equation}
\cos \xi = -\frac{\cos \chi \cos \psi}{\sqrt{\sin^2\chi + \cos^2 \chi \cos^2 \psi}}.
\end{equation}
Given that the bubble has coordinate radius $R$, the intersection of its boundary with this circle (the endpoints of the red line in figure \ref{fig-Vol_calculation}) is simply given by setting $\xi = R$ to find
\begin{equation}
\cos \psi = - \frac{\tan \chi}{\tan R}.
\end{equation}

The area of the boundary of the bubble segment, shown in red in figure \ref{fig-Vol_calculation}, is thus given by
\begin{align}
\mathrm{Area}(R,\chi) &= 4\pi \int_0^{\cos^{-1} \left(-\frac{\tan \chi}{\tan R}\right)} \mathrm{d}\psi\, \sin^2 \psi \nonumber \\
&= 2 \pi \left(\frac{\pi}{2} + \sin^{-1} \left(\frac{\tan \chi}{\tan R}\right) + \frac{\tan \chi}{\tan R} \sqrt{1-\frac{\tan^2 \chi}{\tan^2 R}} \right), \label{area function}
\end{align}
where to get the proper area one must multiply by $\left(\sqrt{\frac{3}{\kappa V}} \sin R \right)^3$.

Meanwhile, the volume of the bubble segment, shown in blue in figure \ref{fig-Vol_calculation}, is given by
\begin{align}
\mathrm{Vol}(R, \chi) &= 4\pi \Bigg\{ \int_0^{\cos^{-1} \left(-\frac{\tan \chi}{\tan R}\right)} \mathrm{d}\psi\, \sin^2 \psi \int_0^R \mathrm{d}\xi\, \sin^3 \xi \nonumber \\
&\qquad\qquad + \int_{\cos^{-1} \left(-\frac{\tan \chi}{\tan R}\right)}^\pi \mathrm{d}\psi\, \sin^2 \psi \int_0^{\cos^{-1} \left( -\frac{\cos \chi \cos \psi}{\sqrt{\sin^2\chi + \cos^2 \chi \cos^2 \psi}} \right)} \mathrm{d}\xi\, \sin^3 \xi \Bigg\} \nonumber \\
&= \mathrm{Area}(R,\chi) \left( \frac{2}{3} - \cos R + \frac{1}{3} \cos^3 R \right) \nonumber \\
&\qquad + \frac{4\pi}{3} \left( \sin^{-1} \left( \frac{\sin\chi}{\sin R}\right) - \sin^{-1} \left( \frac{\tan\chi}{\tan R}\right) + \left( \cos R - 1 \right) \frac{\tan \chi}{\tan R} \sqrt{1- \frac{\tan^2 \chi}{\tan^2 R}}\right) 
, \label{bubble segment volume}
\end{align}
where to get the proper volume one must multiply by $\left(\sqrt{\frac{3}{\kappa V}}\right)^4$.

Note that although $\chi > 0$ was implicitly assumed, one can verify that the expression above satisfies $2\pi^2 \left( \frac{2}{3} - \cos R + \frac{1}{3} \cos^3 R \right) - \mathrm{Vol}(R,-\chi) = \mathrm{Vol}(R, \chi)$, and so \eqref{bubble segment volume} is still valid for $\chi < 0$; similarly, although $R < \frac{\pi}{2}$ was assumed, one has $\frac{4 \pi^2}{3} - \text{Vol}(\pi - R, -\chi) = \text{Vol}(R,\chi)$, and so \eqref{bubble segment volume} is still valid for for $R > \frac{\pi}{2}$.

The extrinsic curvatures of each bubble wall segment give a contribution
\begin{equation}
\left(\sqrt{\frac{3}{\kappa V_f}}\right)^3 \sin^3 R\, \mathrm{Area}(R,\chi) \left( \sqrt{\frac{3 V_f}{\kappa}} \cot R - \sqrt{\frac{3 V_t}{\kappa}} \frac{\sqrt{1- \frac{V_t}{V_f} \sin^2 R}}{\sqrt{\frac{V_t}{V_f}} \sin R} \right),
\end{equation}
which can be absorbed into the volume contribution by defining
\begin{equation}
\mathrm{Vol}'(R, \chi) = \frac{2 \pi^2}{3} - \frac{2}{3} \mathrm{Area}(R,\chi) \cos^3 R + \frac{4 \pi}{3} \left( \sin^{-1} \left( \frac{\sin \chi}{\sin R} \right) + \cos R \frac{\tan \chi}{\tan R} \sqrt{1 - \frac{\tan^2 \chi}{\tan^2 R}} \right), \label{modified bubble segment volume}
\end{equation}
and using this in place of \eqref{bubble segment volume}.

\subsubsection{Consistency conditions on the $\chi$'s}

\begin{figure}
\includegraphics[width=\textwidth]{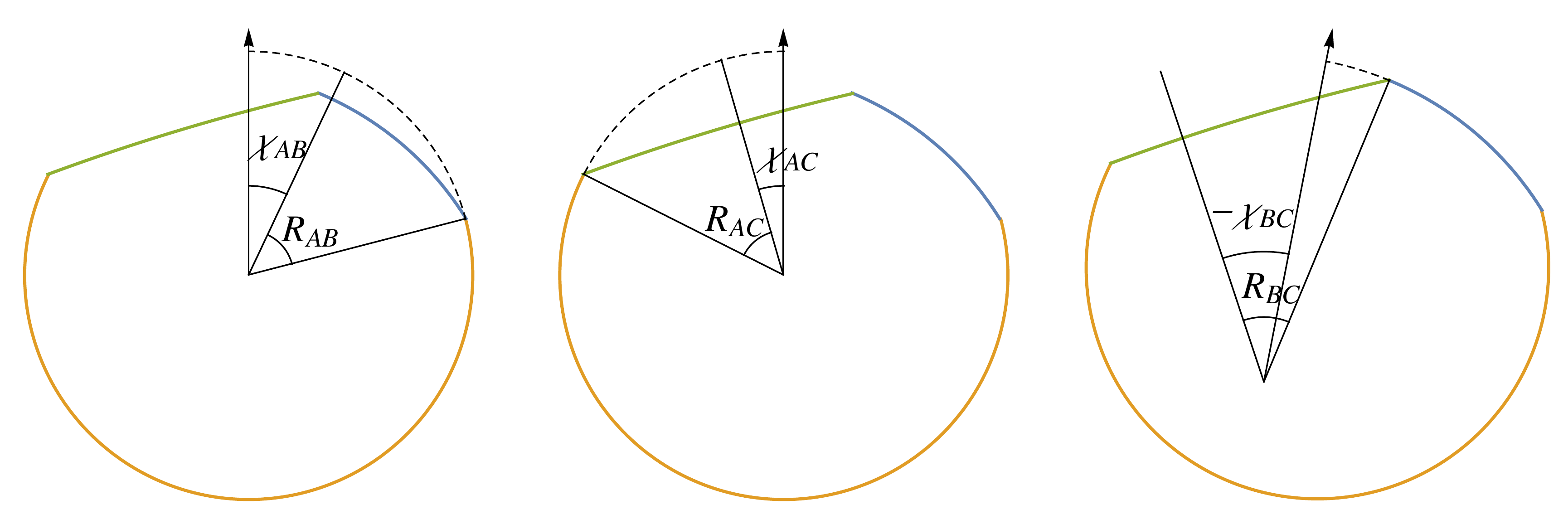}
\caption{`Side-on' view of a barnacle: the yellow segment is the $A$ vacuum, the blue is the $B$ vacuum, and the green is the $C$ vacuum. The dashed lines show, for each false-true interface, the section of false vacuum which has been removed and replaced by a segment of a true vacuum bubble of radius $R$. The $\chi$ parameters control the offsets of the centres of each bubble segment from some coordinate origin, much like the $z$ parameters in the flat space case; they are defined to be positive if there is more than half of that type of bubble. The coordinate origins are chosen such that the `north pole' of the each spherical segment (indicated by the arrow), lies in a plane in the embedding space which contains both the centre of the spherical segment and the junction point.}\label{fig-chis}
\end{figure}

In analogy with the three $z$ parameters in the flat space case, it is convenient to introduce three $\chi$ parameters, alluded to above, which control how the centres of the bubble segments are offset from some coordinate origin. It is convenient to choose these origins, the `north pole' of the each spherical segment, to lie in a plane in the embedding space which contains both the centre of the spherical segment and the junction point. See figure \ref{fig-chis}.

\begin{figure}
\centering
\includegraphics[scale=0.7]{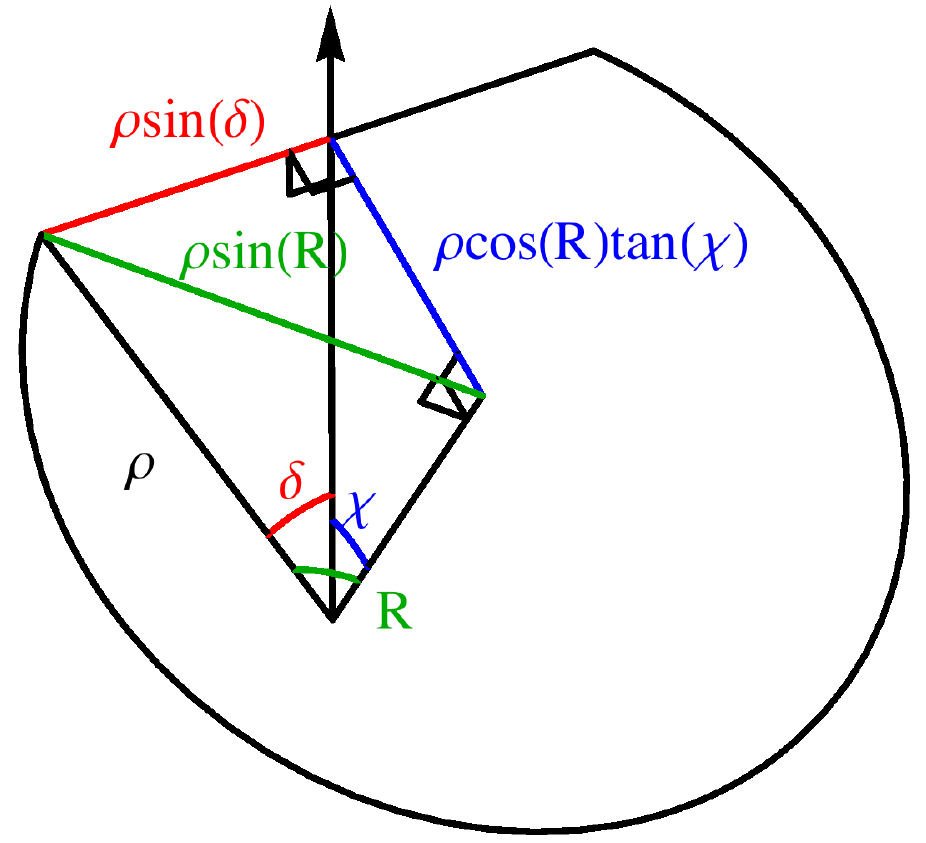}
\caption{The $\chi$ parameters are related to $\delta$, much as the $z$ parameters are related to $r$ in the flat space case (see figure \ref{fig-barnacle_flat_space}). Here $\rho = \sqrt{\frac{3}{\kappa V_A}}$ is the radius of the $A$ vacuum four sphere, and the circle segment is the $AB$ or $AC$ boundary. Since $R_{BC}$ and $\chi_{BC}$ are defined with $B$ as the parent vacuum, but $\delta$ is still defined as in the above figure, for $BC$ one must replace $\sin^2 \delta \to \frac{V_B}{V_A} \sin^2 \delta$.}\label{fig-delta_chi_relation}
\end{figure}

Just as the $z$ parameters are not all independent but satisfy the constraints $R_X^2 - z_X^2 = r^2$, so too must the $\chi$ parameters satisfy a consistency condition relating them all to a single parameter $\delta$, which is a curved space generalisation of the $r$ parameter.
A little geometry---see figure \ref{fig-delta_chi_relation}---reveals that one has\footnote{Note that whilst figure \ref{fig-delta_chi_relation} has $R, \delta < \frac{\pi}{2}$ and $\chi < R$, the relations \eqref{delta def} actually hold generally.}
\begin{equation}
\frac{\cos R_{AB}}{\cos \chi_{AB}} = \frac{\cos R_{AC}}{\cos \chi_{AC}} = \cos \delta \qquad \text{and} \qquad \frac{\cos R_{BC}}{\cos \chi_{BC}} = \sqrt{1 - \frac{V_B}{V_A} \sin^2 \delta}. \label{delta def}
\end{equation}

\subsubsection{Misaligned axes} \label{sec-theta}

\begin{figure}
\centering
\includegraphics[scale=0.5]{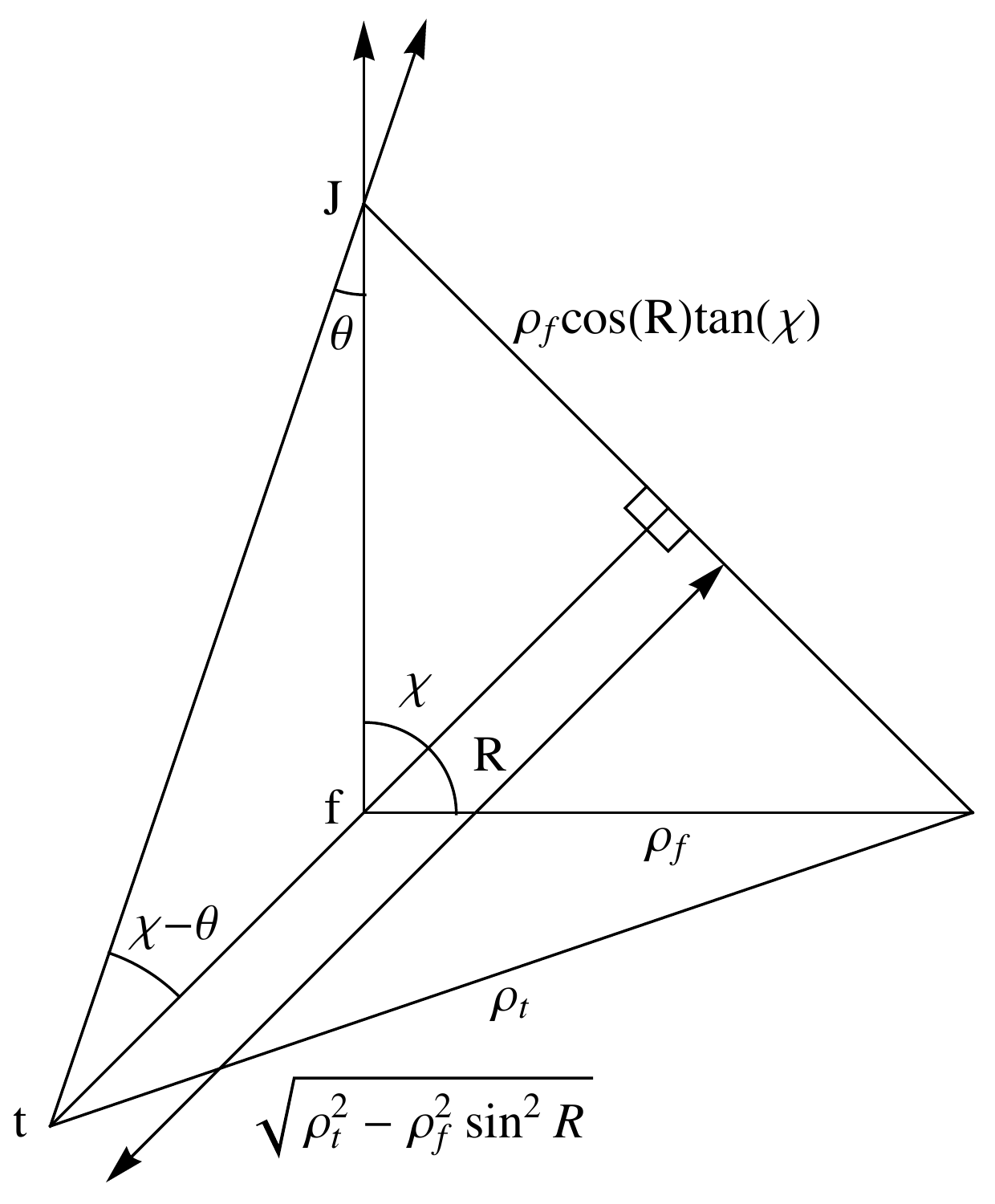}
\caption{The relationship between the misalignment angle $\theta$ and the other parameters. $\rho_i = \sqrt{\frac{3}{\kappa V_i}}$ is the radius of the four sphere; $J$ is the junction point where all three vacua meet, $f$/$t$ is the centre of the false/true vacuum four sphere (for the pair of vacua being considered).}\label{fig-theta_relation}
\end{figure}

Since the radii of the four spheres which make up the various bubble segments are not all equal, the planes in the embedding space which go through the junction point and each of the centres of the spheres are misaligned. From figure \ref{fig-theta_relation} one sees that these misalignment angles $\theta$ satisfy
\begin{equation}
\tan \left( \chi - \theta \right) = \sqrt{\frac{V_t}{V_f}} \frac{\cos R \tan \chi}{\sqrt{1 - \frac{V_t}{V_f} \sin^2 R}}. \label{theta def}
\end{equation}

One also sees that when dealing with quantities on the true vacuum side, one must make the replacements:
\begin{equation}
\sin R \to \sqrt{\frac{V_t}{V_f}} \sin R \qquad \text{and} \qquad \chi \to \chi - \theta.
\end{equation}

\subsubsection{Conical singularity at the junction}

The final ingredient to consider is the fact that the non-zero energy $\mu$ possessed by the junction two sphere induces a conical singularity at its location in the remaining co-dimension two $(\xi, \psi)$ space. One can calculate this by transforming to a coordinate system in each segment of the barnacle which near the junction point looks like flat space in polar coordinates (crossed with a two sphere), and then determining the total range of the polar angular coordinate. 

The details of the calculation are relegated to appendix \ref{app-deficit angle calculation}, and here I just quote the result for the deficit angle:
\begin{align}
\Delta = &\cot^{-1} \left(\frac{\tan (\chi_{AB} - \theta_{AB})}{ \sqrt{\frac{V_B}{V_A}} \sin \delta} \right) - \cot^{-1} \left( \frac{\tan \chi_{AB}}{\sin \delta} \right) \nonumber \\
&+ \cot^{-1} \left(\frac{\tan (\chi_{AC} - \theta_{AC})}{ \sqrt{\frac{V_C}{V_A}} \sin \delta} \right) - \cot^{-1} \left( \frac{\tan \chi_{AC}}{\sin \delta} \right) \nonumber \\
&+ \cot^{-1} \left( \frac{\tan (\chi_{BC} - \theta_{BC})}{\sqrt{\frac{V_C}{V_A}} \sin \delta} \right) - \cot^{-1} \left( \frac{\tan \chi_{BC}}{\sqrt{\frac{V_B}{V_A}} \sin \delta} \right), \label{deficit angle}
\end{align}
where the range of the inverse cotangent is taken to be $(0,\pi)$.

It is perhaps interesting to note that the condition for the barnacle geometry to be continuous when embedded in $\mathbb{R}^5$, which one can determine to be
\begin{equation}
\theta_{AB} + \theta_{AC} + \theta_{BC} = 0,
\end{equation}
is not equivalent to the vanishing of the deficit angle.

\subsubsection{Flat space limit}

A useful sanity check one can perform at this stage is to compare the flat space limit of \eqref{Sbarn with grav} with \eqref{Sbarn flat space}. The flat space limit can be accessed by writing\footnote{Note that $R_{ft} = \sqrt{\frac{\kappa V_f}{3}} R^\text{flat}_{ft}$ does not agree with putting $\frac{3\sigma_{ft}}{\epsilon_{ft}} = R^\text{flat}_{ft}$ in the curved space solution for the bubble radius \eqref{R solution}, except at first order in $\kappa$, but it is only the lowest order in $\kappa$ which interests us here.}
\begin{equation}
R_{ft} \to \sqrt{\frac{\kappa V_f}{3}} R_{ft}, \qquad \chi_{ft} \to \sqrt{\frac{\kappa V_f}{3}} z_{ft}, \qquad \delta \to \sqrt{\frac{\kappa V_A}{3}} r,
\end{equation}
and then taking $\kappa \to 0$.
One finds
\begin{align}
&\lim_{\kappa \to 0} \left[ \sqrt{\frac{3}{\kappa V}} \sin\left(\sqrt{\frac{\kappa V}{3}} R\right) \right]^3 \mathrm{Area}\left(\sqrt{\frac{\kappa V}{3}} R,\sqrt{\frac{\kappa V}{3}} z\right) \nonumber \\
&\qquad = 2\pi R^3 \left( \frac{\pi}{2} + \sin^{-1}\left(\frac{z}{R}\right) + \frac{z}{R}\sqrt{1- \left(\frac{z}{R}\right)^2} \right) \\
&\lim_{\kappa \to 0} \left[ \sqrt{\frac{3}{\kappa V}} \right]^4 \mathrm{Vol}\left(\sqrt{\frac{\kappa V}{3}} R,\sqrt{\frac{\kappa V}{3}} z\right) = \lim_{\kappa \to 0} \left[ \sqrt{\frac{3}{\kappa V_t}} \right]^4 \mathrm{Vol}\left(\sqrt{\frac{\kappa V_f}{3}} R,\sqrt{\frac{\kappa V_f}{3}} z - \theta \right) \nonumber \\
&\qquad = \frac{\pi}{2} R^4 \left( \frac{\pi}{2} + \sin^{-1}\left(\frac{z}{R}\right) + \frac{1}{3}\frac{z}{R}\left( 5 - 2\left(\frac{z}{R}\right)^2 \right)\sqrt{1- \left(\frac{z}{R}\right)^2}\right),
\end{align}
from which one can verify that the terms in \eqref{Sbarn with grav} which \emph{do not} involve $\mathcal{R}$, $\mathcal{K}$, or $\Delta$, when taken with $- \frac{24 \pi^2}{\kappa^2 V_A}$  reproduce exactly \eqref{Sbarn flat space}. 
Meanwhile, using
\begin{equation}
\lim_{\kappa \to 0} \left[ \sqrt{\frac{3}{\kappa V}} \sin\left(\sqrt{\frac{\kappa V}{3}} r \right) \right]^2 \frac{\Delta}{\kappa} = \frac{r^3}{6} \left( (V_A-V_B) z_{AB} + (V_A - V_C) z_{AC} + (V_B - V_C) z_{BC} \right),
\end{equation}
one can determine the the remaining terms, \ie those involving $\mathcal{R}$, $\mathcal{K}$, or $\Delta$, along with the term $2\frac{24 \pi^2}{\kappa^2 V_A}$, cancel among themselves.

\subsection{Extremising the Action}

Given the scalar field potential (and hence the parameters $V_{A,B,C}$, $\sigma_{AB,AC,BC}$, and $\mu$) the action is then a function of four variables: $R_{AB}$, $R_{AC}$, $R_{BC}$, and $\delta$. In order to reduce the expressions for the various contributions to the action to functions of just these variables, several relations are useful.

The deficit angle, \eqref{deficit angle}, only depends on
\begin{align}
\frac{\tan \chi_{ft}}{\sin \delta} &= \mathrm{sign}(\chi_{ft})\mathrm{sign}(\cos \delta) \frac{\sqrt{\frac{\sin^2 R_{ft}}{\frac{V_f}{V_A} \sin^2 \delta} - 1}}{\cos R_{ft}} \\
\frac{\tan (\chi_{ft} - \theta_{ft})}{ \sqrt{\frac{V_t}{V_A}}\sin \delta} &= \mathrm{sign}(\chi_{ft})\mathrm{sign}(\cos \delta) \frac{\sqrt{\frac{\sin^2 R_{ft}}{\frac{V_f}{V_A} \sin^2 \delta} - 1}}{\sqrt{1 - \frac{V_t}{V_f} \sin^2 R_{ft}}}.
\end{align}

Meanwhile the area of a bubble wall segment, \eqref{area function}, just depends on
\begin{equation}
\frac{\tan \chi_{ft}}{\tan R_{ft}} = \frac{\tan (\chi_{ft} - \theta_{ft})}{\tan \left[ \sin^{-1} \left( \sqrt{\frac{V_t}{V_f}} \sin R_{ft} \right) \right]} = \mathrm{sign}(\chi_{ft})\mathrm{sign}(\cos \delta) \sqrt{1 - \frac{V_f}{V_A} \frac{\sin^2 \delta}{\sin^2 R_{ft}}}. \label{tan chi over tan R}
\end{equation}
Note that the first equality means that, as one would expect, the area of the boundary is the same as measured on either side of it.

Finally, the volume of a bubble segment, \eqref{modified bubble segment volume}, depends additionally on
\begin{align}
\frac{\sin \chi_{ft}}{\sin R_{ft}} &= \mathrm{sign}(\chi_{ft}) \sqrt{\frac{1- \frac{V_f}{V_A}\frac{\sin^2 \delta}{\sin^2 R_{ij}}}{1- \frac{V_f}{V_A}\sin^2 \delta}} \\
\frac{\sin (\chi_{ft} - \theta_{ft})}{ \sqrt{\frac{V_t}{V_f}}\sin R_{ft}} &= \mathrm{sign}(\chi_{ft}) \sqrt{\frac{1- \frac{V_f}{V_A}\frac{\sin^2 \delta}{\sin^2 R_{ij}}}{1- \frac{V_t}{V_A}\sin^2 \delta}}.
\end{align}
Crucially, note that these quantities do not depend on $\mathrm{sign}(\cos \delta)$.

One could directly extremise the action at this stage, however it proves beneficial to first consider achieving this minimisation by applying Einstein's equations in the form of the Israel junction conditions at the bubble walls, along with relating the conical deficit angle to $\mu$.

\subsubsection{Israel Junction Conditions at the Bubble Walls}

The Israel junction conditions \cite{Israel:1966rt} allow one to relate the discontinuity in extrinsic curvature on either side of the bubble wall to the tension of the that wall. Taking care that here the extrinsic curvature should be calculated using a normal pointing into the false vacuum on both sides of the wall (whereas in the action the normal points away from whichever region is being considered), one has
\begin{equation}
\mathcal{K}_f - \mathcal{K}_t = \sqrt{3 \kappa V_f} \frac{\cos R}{\sin R} \mp \sqrt{3 \kappa V_t} \frac{\sqrt{1 - \frac{V_t}{V_f} \sin^2 R}}{\sqrt{\frac{V_t}{V_f}} \sin R} = -\frac{3}{2} \kappa \sigma, \label{Israel}
\end{equation}
where the upper (lower) sign refers the to the case where the true vacuum region is less (more) than half a four sphere. One sees that only the upper sign allows for the wall tension to be positive, \ie the region of lower vacuum energy must be less than half a four sphere, whilst the false vacuum region is at this point not so constrained.

Equation \eqref{Israel} can be solved to give
\begin{equation}
\frac{3}{\kappa V_f} \sin^2 R = \left( \frac{\kappa V_f}{3}  + \left[ \frac{V_f-V_t}{3 \sigma} - \frac{\kappa \sigma}{4} \right]^2 \right)^{-1}. \label{R solution}
\end{equation}
This holds regardless of whether the wall is a sphere or just a segment, and so it means that each bubble segment in a barnacle has the same radius it would would have if there were just a single spherical bubble.

\subsubsection{Conical Singularity}

The angular deficit at the junction point is related to the energy at that point by \cite{Vilenkin:1981zs}
\begin{equation}
\mu = \frac{\Delta}{\kappa}. \label{conical singularity einstein}
\end{equation}
For $\mu < 0$ this means that $\Delta$ actually describes \emph{surfeit} angle. Given that \eqref{R solution} fixes the three $R$'s, this condition then fixes $\delta$, giving the barnacle geometry which is a solution to the Euclidean Einstein equations, up to some caveats mentioned in the next subsection.

It is interesting to note from \eqref{Sbarn with grav} that when evaluated on shell, the deficit angle and $\mu$ cancel and drop out of the action (though they of course affect the location of the extremum).

\subsubsection{Directly Extremising the Action}

It is also possible to extremise the action by directly finding its stationary points as a function of $\{R_X, \delta\}$, as is done in flat space. It is a useful sanity check that this gives the same results as the previous subsections---up to some important caveats. One has
\begin{align}
\frac{\partial S_b}{\partial R_X} =& \frac{6\pi^2}{\kappa^2V_f} \sin^2 R_X \cos R_X
\left[\frac{\sigma}{2} \sqrt{\frac{3\kappa}{V_f}} + \cot R_X - \frac{\sqrt{1 -  \frac{V_t}{V_f} \sin^2 R_X}}{\sin R_X} \right] \nonumber \\
& \times \Bigg\{ 3\pi + 2\mathrm{sign}(\chi_X) \mathrm{sign}(\cos \delta) \Bigg( 3 \sin^{-1} \sqrt{1 - \frac{\sin^2 \delta}{\sin^2 R_X}} + 3 \frac{\sin\delta}{\sin R_X} \sqrt{1 - \frac{\sin^2 \delta}{\sin^2 R_X}} \nonumber \\
& \hspace{3.2in} + 2 \frac{\sin^3\delta}{\sin^3 R_X} \left(1 - \frac{\sin^2 \delta}{\sin^2 R_X}\right)^{-\frac{1}{2}} \Bigg) \Bigg\} \nonumber \\
&+ \frac{12 \pi}{\kappa^2 V_f} \mathrm{sign}(\chi_X) \frac{\sin \delta \cos R_X}{\sin^2 R_X} \left(1 - \frac{\sin^2 \delta}{\sin^2 R_X}\right)^{-\frac{1}{2}} \nonumber \\
& \hspace{0.5in}\times \left( \frac{\mathrm{sign}(\cos R_X) - \mathrm{sign}(\cos \delta)}{\cos R_X} + \frac{\mathrm{sign}(\cos\delta) - 1}{\frac{V_t}{V_f} \sqrt{1 - \frac{V_t}{V_f} \sin^2 R_X}} \right). \label{dSdR}
\end{align}
The expression in the square brackets is just that which vanishes when the Israel junction conditions are satisfied---see \eqref{Israel}. For the derivative with respect to $\delta$ one has
\begin{align}
\frac{\partial S_b}{\partial \delta} = &\frac{24\pi}{\kappa^2 V_A} \sin\delta\cos\delta \left( \kappa \mu - \Delta \right) \nonumber \\
&- \sum_X \frac{12\pi}{\kappa^2 V_A} \mathrm{sign}(\chi_X) \frac{\cos \delta}{\sin R_X} \left(1 - \frac{V_f}{V_A} \frac{\sin^2 \delta}{\sin^2 R_X} \right)^{-\frac{1}{2}} \nonumber \\
&\qquad \times \Bigg\{ 2\mathrm{sign}(\cos \delta) \sin^2\delta \sin R_X \left[\frac{\sigma}{2} \sqrt{\frac{3\kappa}{V_f}} + \cot R_X - \frac{\sqrt{1 - \frac{V_t}{V_f} \sin^2 R_X}}{\sin R_X} \right] \nonumber \\
&\qquad + \frac{\cos R_X}{\cos^2 \delta} \left( \mathrm{sign}(\cos R_X) - \mathrm{sign}(\cos \delta) \right) + \frac{\sqrt{1 - \frac{V_t}{V_f} \sin^2 R_X}}{1 - \frac{V_t}{V_f} \sin^2 \delta} \left( \mathrm{sign}(\cos \delta) - 1 \right) \Bigg\}, \label{dSddelta}
\end{align}
and again one notices on the first line the quantity which vanishes when Einstein's equations at the conical singularity are satisfied---see \eqref{conical singularity einstein}.

When $\delta < \frac{\pi}{2}$ and $R_X < \frac{\pi}{2}$ for all $X$, the final line of each of \eqref{dSdR} and \eqref{dSddelta} vanishes, and the action is indeed made stationary when the junction conditions are satisfied. On the other hand, when one (or more) of the bubble radii are larger than $\frac{\pi}{2}$ then these extra terms mean that the action is no longer stationary when the junction conditions are satisfied, and in fact it may be that there is no solution which makes the action stationary. I take this to mean that barnacle geometries can only exist when the bubbles are sufficiently small (\viz $R < \frac{\pi}{2}$); this may seem surprising, since barnacles can always exist in flat space, however it is precisely for larger bubbles that the curved (and finite) nature of (Euclidean) de Sitter space makes itself felt most keenly. 

Thus, in terms of vacuum energies and wall tensions, barnacles can only exist if
\begin{equation}
V_f - V_t > \frac{3}{4}\kappa \sigma_{ft}^2, \label{<1/2 four sphere bound}
\end{equation}
is satisfied for each pair of vacua. If these conditions are satisfied, then, much like the flat space barnacle, for $\mu = 0$ a solution always exists, whereas this is not always true for non-zero $\mu$, as will be discussed in section \ref{sec-negative modes}.

It is interesting to note the similarity of the bound \eqref{<1/2 four sphere bound} to the gravitational quenching which arises when considering spherical bubble nucleation in a Minkowski or anti-de Sitter false vacuum: if $\left(\sqrt{-V_f} - \sqrt{-V_t}\right)^2 > \frac{3}{4}\kappa \sigma_{ft}^2$ is not satisfied, then the false vacuum is rendered stable \cite{Parke:1982pm}. This occurs because, if the difference between the vacuum energies is too small, then conservation of (pseudo-)energy would require the resulting bubble to have infinite size, whereas the lack of a corresponding conserved quantity means that no such quenching occurs in decays from de Sitter vacua, as are being considered in this paper. Nonetheless this explains the similarity with \eqref{<1/2 four sphere bound}, since the latter arises because barnacle actions cannot be made stationary if the radius of the bubble is too large, and this maximum comes from setting to zero the quantity in square brackets in \eqref{R solution}, whilst the quenching bound can be derived by setting the quantity in the round brackets in \eqref{R solution} to zero.

Finally, although I am focussing on de Sitter vacua, let me briefly discuss the fate, when any of the vacuum energies go negative, of the bound \eqref{<1/2 four sphere bound}, since it is stronger than the quenching bound. When the false vacuum is de Sitter, regardless of the nature of the true vacuum, there will still be present terms in $\frac{\partial S_b}{\partial R_X}$ and $\frac{\partial S_b}{\partial \delta}$ which lead to the bound \eqref{<1/2 four sphere bound}; on the other hand when both vacua are anti-de Sitter, the results get modified simply by $R_X \to i R_X$, and hence, for that pair of vacua, the bound \eqref{<1/2 four sphere bound} would disappear and be replaced by the quenching bound.\footnote{Similarly, when $V_A$ turns negative one must make the replacement $\delta \to i \delta$.} 

\subsection{Expression for the Action}

The result of applying the previous subsections to the action \eqref{Sbarn with grav} yields the following expression for the action of a barnacle instanton:
\begin{align}
S_b = & - \left(\frac{3}{\kappa}\right)^2 \Bigg[ \frac{1}{V_A} \left\{ - \mathrm{Vol}'(R_{AB},\chi_{AB}) - \mathrm{Vol}'(R_{AC},\chi_{AC}) \right\} \nonumber \\
&\qquad + \frac{1}{V_B} \left\{ \mathrm{Vol}'\left(\tilde{R}_{AB}, \chi_{AB} - \theta_{AB} \right) - \mathrm{Vol}'(R_{BC}, \chi_{BC}) \right\} \nonumber \\
&\qquad + \frac{1}{V_C} \left\{ \mathrm{Vol}'\left(\tilde{R}_{AC}, \chi_{AC} - \theta_{AC} \right) + \mathrm{Vol}'\left(\tilde{R}_{BC}, \chi_{BC} - \theta_{BC} \right) \right\} \Bigg] \nonumber \\
&+ \left(\frac{3}{\kappa}\right)^{3/2} \Bigg( \sigma_{AB} \frac{\sin^3 R_{AB}}{V_A^{3/2}} \mathrm{Area}(R_{AB},\chi_{AB}) + \sigma_{AC} \frac{\sin^3 R_{AC}}{V_A^{3/2}} \mathrm{Area}(R_{AC},\chi_{AC}) \nonumber \\
& \qquad\qquad\qquad\qquad + \sigma_{BC} \frac{\sin^3 R_{BC}}{V_B^{3/2}} \mathrm{Area}(R_{BC},\chi_{BC}) \Bigg), \label{Sbarn with grav - evaluate}
\end{align}
where $\tilde{R}_{ft} = \sin^{-1} \left( \sqrt{\frac{V_t}{V_f}} \sin R_{ft} \right)$, and since the region of lower vacuum energy must be less than half of a four sphere, one does not need to worry about the branch of the inverse sine; the $\mathrm{Vol}'$ and $\mathrm{Area}$ functions are given by \eqref{modified bubble segment volume} and \eqref{area function} respectively.

\subsubsection{Signs of the $\chi$ Parameters}

A priori there are eight possible sets of signs of $\chi$, and, as in the flat space case, when $\mu = 0$ the solution of \eqref{conical singularity einstein} will pick out two, related by an overall sign change, only one of which is a valid barnacle.

Two of these sets---\viz $\{\mathrm{sign}(\chi_{AB}), \mathrm{sign}(\chi_{AC}), \mathrm{sign}(\chi_{BC})\} = \{\pm, \pm, \pm \}$---correspond to having either more than half of each type of bubble or less than half of each type of bubble. They can be eliminated as possibilities in the flat space case as they violate energy conservation (there being respectively either too little or too much wall for its tension offset the energy density of the interiors), however as there is no global notion of conserved energy in de Sitter space this cannot be used to eliminate them here. On the other hand, each of the three lines in \eqref{deficit angle} is a positive quantity multiplied by $\mathrm{sign}(\chi_X)$, and so $\{\pm, \pm, \pm \}$ can only be a solution for $\mu \gtrless 0$. In fact, it will turn out in the next section that these signs always lead to a barnacle with too many or too few negative modes.

Another restriction one can place is that the $BC$ bubble segment should be able to fit inside the $AB$ bubble segment. From \eqref{B range} one sees that this is equivalent to the condition
\begin{equation}
\chi_{AB} - \theta_{AB} > \chi_{BC}. \label{room for B}
\end{equation}
Given that $\mathrm{sign}(\chi_{AB} - \theta_{AB}) = \mathrm{sign}(\chi_{AB})$, this immediately rules out $\{ -, \pm, + \}$. Using \eqref{tan chi over tan R} one can show that \eqref{room for B} implies $\sin^2 \delta > 1$, and hence the impossibility of a solution, when
\begin{align}
\frac{1}{V_A} \sin^2 R_{AB} \lessgtr \frac{1}{V_B} \sin^2 R_{BC} &\qquad \text{ for } \mathrm{sign}(\chi_{AB}) = \mathrm{sign}(\chi_{BC}) = \pm \nonumber \\
\implies \frac{V_A - V_B}{3 \sigma_{AB}} + \frac{\kappa \sigma_{AB}}{4} \gtrless \frac{V_B - V_C}{3 \sigma_{BC}} - \frac{\kappa \sigma_{BC}}{4} &\qquad \text{ for } \mathrm{sign}(\chi_{AB}) = \mathrm{sign}(\chi_{BC}) = \pm.
\end{align}

With these considerations, when $\mu = 0$, the solution of \eqref{conical singularity einstein} will thus yield a unique set of signs of the $\chi$ parameters. 

\subsubsection{Negative Modes} \label{sec-negative modes}

As explained in section \ref{sec-multi neg}, barnacles describe two decay processes (the production of the initial bubble, and the subsequent decay of its wall) and so there should be exactly two negative modes of fluctuations about them.\footnote{One negative mode would describe a barnacle which cannot be produced in two stages, but can only be produced `fully formed,' which does not seem realistic, whilst three, or more, negative modes would seem to require some unknown intermediate process, which again does not seem realistic.}

Considering the derivative of the action with respect to $R_X$, \eqref{dSdR}, one has that $\frac{\partial^2 S_b}{\partial R_X \partial \delta}$ vanishes on the barnacle solution. Therefore we need only to consider the diagonal elements of the Hessian of $S_b$.

In \eqref{dSdR} the derivative with respect to $R_X$ of the quantity in square brackets is always negative at the value which solves \eqref{Israel}; this, along with the fact that the sign of the expression in curly brackets is equal to $\mathrm{sign}(\chi_X)$, means that on a barnacle solution one has\footnote{Recall that barnacles can only have $\cos R_X > 0$ and $\cos \delta > 0$.}
\begin{equation}
\mathrm{sign} \left( \frac{\partial^2 S_b}{\partial R_X^2} \right) = - \mathrm{sign}\left(\chi_X \right). \label{d2SdR2}
\end{equation}
Incidentally, this means that the $\{\mathrm{sign}(\chi_{AB}), \mathrm{sign}(\chi_{AC}), \mathrm{sign}(\chi_{BC})\} = \{\pm, \pm, \pm\}$ possibilities can be discarded since they could never lead to a barnacle with exactly two negative modes.

Rather than considering $\frac{\partial^2 S_b}{\partial \delta^2}$ directly, when $\mu = 0$ it is simpler to use the fact that there is only one non-trivial solution of \eqref{conical singularity einstein}, and consider the small $\delta$ behaviour of $S_b$; in particular expanding the first line of \eqref{dSddelta} yields
\begin{equation}
\frac{\partial S_b}{\partial \delta} = \left(\frac{3}{\kappa V_A}\right) 8\pi \mu \delta - 4\pi \delta^2 \left(\frac{3}{\kappa V_A}\right)^\frac{3}{2}\sum_X \mathrm{sign}(\chi_X) \sigma_X + \mathcal{O}(\delta^3) \label{dSddelta small delta}
\end{equation}
Therefore, given that all the wall tensions are positive, for $\mu = 0$, on the barnacle solution one has
\begin{equation}
\frac{\partial^2 S_b}{\partial \delta^2}
\begin{cases}
> 0 & \text{ for two }\chi\text{'s positive,} \\
< 0 & \text{ for two }\chi\text{'s negative}.
\end{cases}
\end{equation}

\begin{figure}[tp]
\begin{subfigure}{0.5\textwidth}
\includegraphics[width=\textwidth]{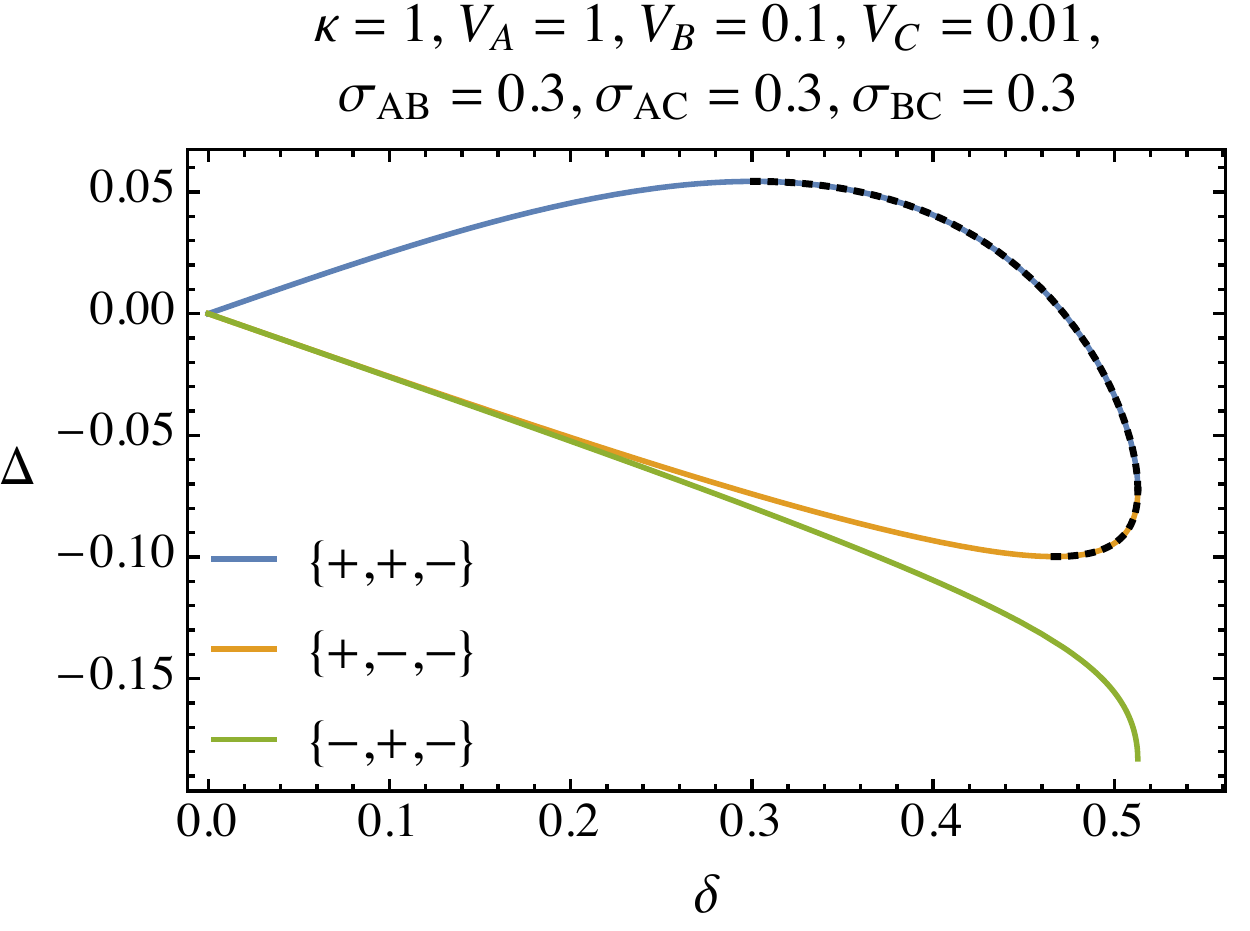}
\end{subfigure}
\hfill
\begin{subfigure}{0.5\textwidth}
\includegraphics[width=\textwidth]{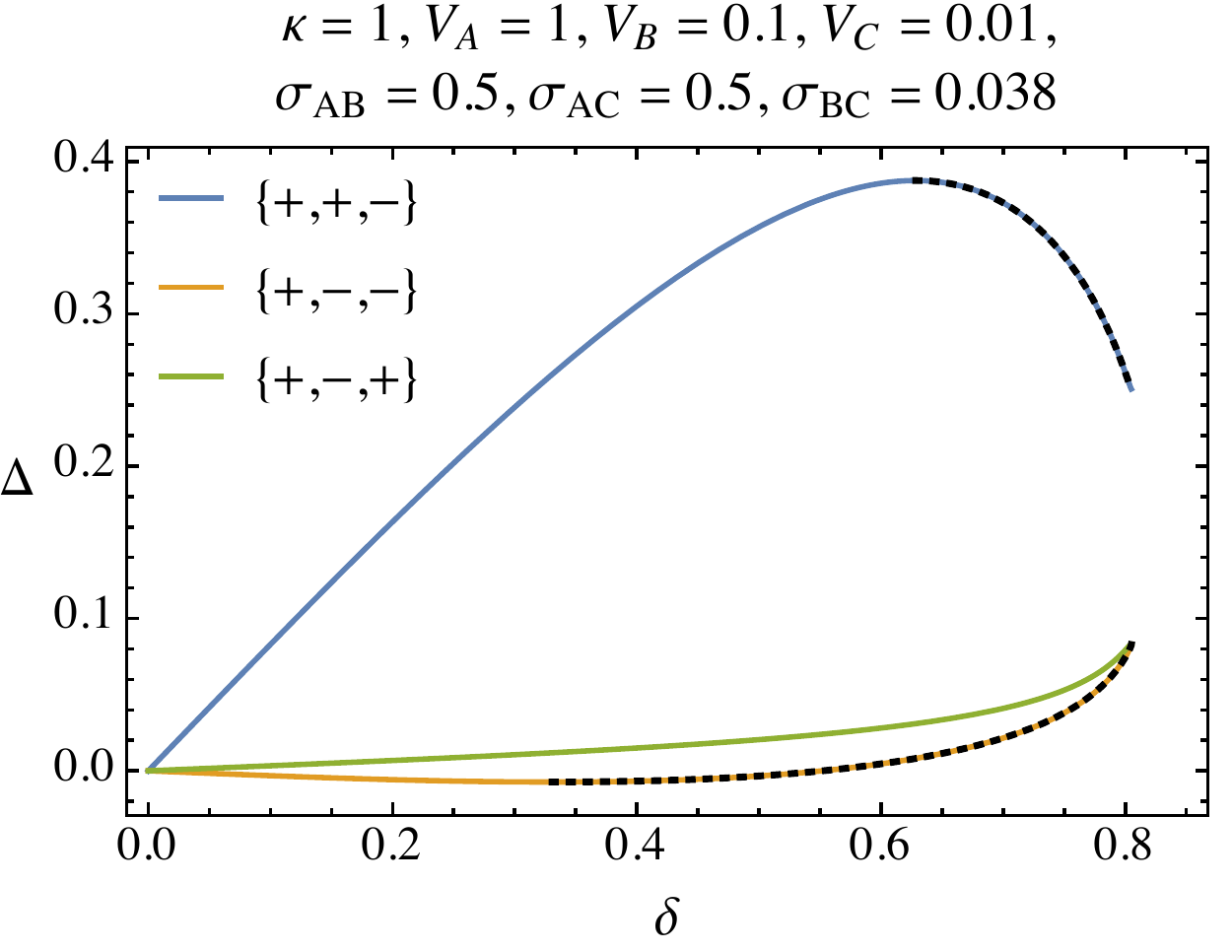}
\end{subfigure}
\caption{The deficit angle $\Delta$ as a function of the junction point location $\delta$, for two different sets of vacuum energies and wall tensions, and for various allowed signs of the $\chi$ parameters. The black dotted lines show the range in which barnacles exist, \ie when $\kappa \mu = \Delta$ has a solution \emph{and} $\frac{\partial \Delta}{\partial \delta}$ has the correct sign; $\{+,+,-\}$ and $\{+,-,+\}$ mean that there are two negative modes due to $R_X$ fluctuations, and so the slope of $\Delta$ must be negative in order that $\delta$ fluctuations give a positive mode; this is reversed for $\{+,-,-\}$ and $\{-,+,-\}$.}\label{fig-Delta_delta}
\end{figure}

When $\mu \neq 0$ determining the sign of $\frac{\partial^2 S_b}{\partial \delta^2}$ becomes more complicated. From \eqref{dSddelta} one has that it is given by minus the sign of $\frac{\partial \Delta}{\partial \delta}$. Therefore, given a particular set of wall tensions and vacuum energies and a plot of $\Delta(\delta)$---for example see figure \ref{fig-Delta_delta}---one can determine the allowed signs of $\chi_X$ based on the derivative of the curve. 

One negative mode must come from one of the $R_X$, but it is worth pointing out that there are two qualitatively different types of barnacle, depending on the origin of the second negative mode. If it comes from one of the $R_X$, then from \eqref{d2SdR2} one sees that both $\chi_{AB}$ and one of $\chi_{AC}$ and $\chi_{BC}$ will be positive, and hence there is more than half of an $AB$ bubble \emph{and} more than half of an $AC$ or $BC$ bubble. On the other hand, if the second negative mode comes from $\delta$, then only one $\chi$ will be positive, and hence there will only be more than half of \emph{either} an $AB$ \emph{or} an $AC$ bubble, and less than half of the other (and less than half of $BC$).

\section{Comparative Decay Rates} \label{sec-comp decay rates}

Given an expression for the barnacle action, let us now compare it with actions for spherical bubbles. 
Consider an $AB$ and an $AC$ bubble which are tangent; this can be thought of as a barnacle with $\delta = 0$ and $\chi_{AB} = R_{AB}$, $\chi_{AC} = R_{AC}$, and $\chi_{BC} = -R_{BC}$. From the small $\delta$ expansion of the action, \eqref{dSddelta small delta}, one has in this case, for $\mu = 0$:
\begin{equation}
\frac{\partial S_b}{\partial \delta} = - 4\pi \delta^2 \left(\frac{3}{\kappa V_A}\right)^\frac{3}{2} (\sigma_{AB} + \sigma_{AC} - \sigma_{BC}) + \mathcal{O}(\delta^4),
\end{equation}
which is negative, due to the triangle inequalities for the wall tensions, \eqref{triangle inequality}. Therefore merging the bubbles to form a barnacle \emph{decreases} the action. Similarly one can consider a $BC$ bubble tangent to the wall of an $AB$ bubble to again find that, due to the wall tension triangle inequalities, increasing $\delta$ and forming a barnacle by moving the $C$ region slightly outside of the $B$ region decreases the action. 

On the other hand, a solitary $AB/C$ bubble is effectively a barnacle with $\delta = 0$, $\chi_{AB/C} = R_{AB/C}$, $\chi_{AC/B} = -R_{AC/B}$, and $\chi_{BC} = -R_{BC}$, and by similar arguments one finds that increasing $\delta$ will increase the action.

Thus for $\mu = 0$ one has
\begin{equation}
S_{AB \text{ or } AC} < S_b < S_{AB} + S_{AC \text{ or } BC}. \label{barnacle action inequalities}
\end{equation}

Physically, these inequalities can be understood in the following way: in merging an $AB$ and an $AC$ bubble, one increases the volume which is in the false vacuum, but this is outweighed by the reduction in the total area of the walls; alternatively the triangle inequalities tell us that it is easer to go from $B$ to $C$ directly, rather than via $A$, and so given the initial $AB$ and $AC$ bubbles, the action can be reduced by including a bit of $BC$ wall.

In pulling a $BC$ bubble out of an $AB$ bubble one is increasing the area of the walls, but also increasing the volume in the intermediate and true vacua, which means that the action decreases. Again one can also think of this as due to triangle inequalities implying that, given a region in the $C$ vacuum, the action can be reduced by introducing an $AC$ interface.

Finally, replacing the section of the wall of an $AB/C$ bubble with a barnacle impedes the direct interface between $A$ and $B/C$, and hence increases the action. Or, in other words, more walls are created, but this is not outweighed by the reduction in volume of the false vacuum region.

As explained in section \ref{sec-multi neg}, a barnacle mediates the decay of the wall of a bubble, and so in the limit that the decay rate is insensitive to the pre-factor to the exponential, the right hand of the above inequalities are telling us that, given a bubble, it is more likely that a section of its wall decays, than that a spherical bubble of the third type of vacuum is produced (either inside or outside the original bubble).\footnote{The left hand inequalities in \eqref{barnacle action inequalities} are just expressing the uncontroversial statement that it is more likely for a bubble to appear, than for a bubble to appear \emph{and} have its wall decay.}

All that remains is to consider the production of a bubble of the \emph{same} type as the initial bubble. If the rate of production of these bubbles is less than the rate of production of the other type, then clearly barnacles will be favoured. On the other hand, if the rate of production of these bubbles is vastly greater than the rate of production of the other type, then it seems unlikely that barnacles could ever be competitive. This leaves a window---when the rate of production of the two types of bubble are comparable---in which the simple arguments above are not sufficient to determine the favourableness of barnacles.

\subsection{Barnacle Action Results}

In this section I will present some representative results of the value of the barnacle action, and its relation to various spherical bubble processes. Throughout I will set the value of the reduced Planck mass to one, \ie $\kappa = 1$.

\begin{figure}[tp]
\centering
\includegraphics[scale=0.7]{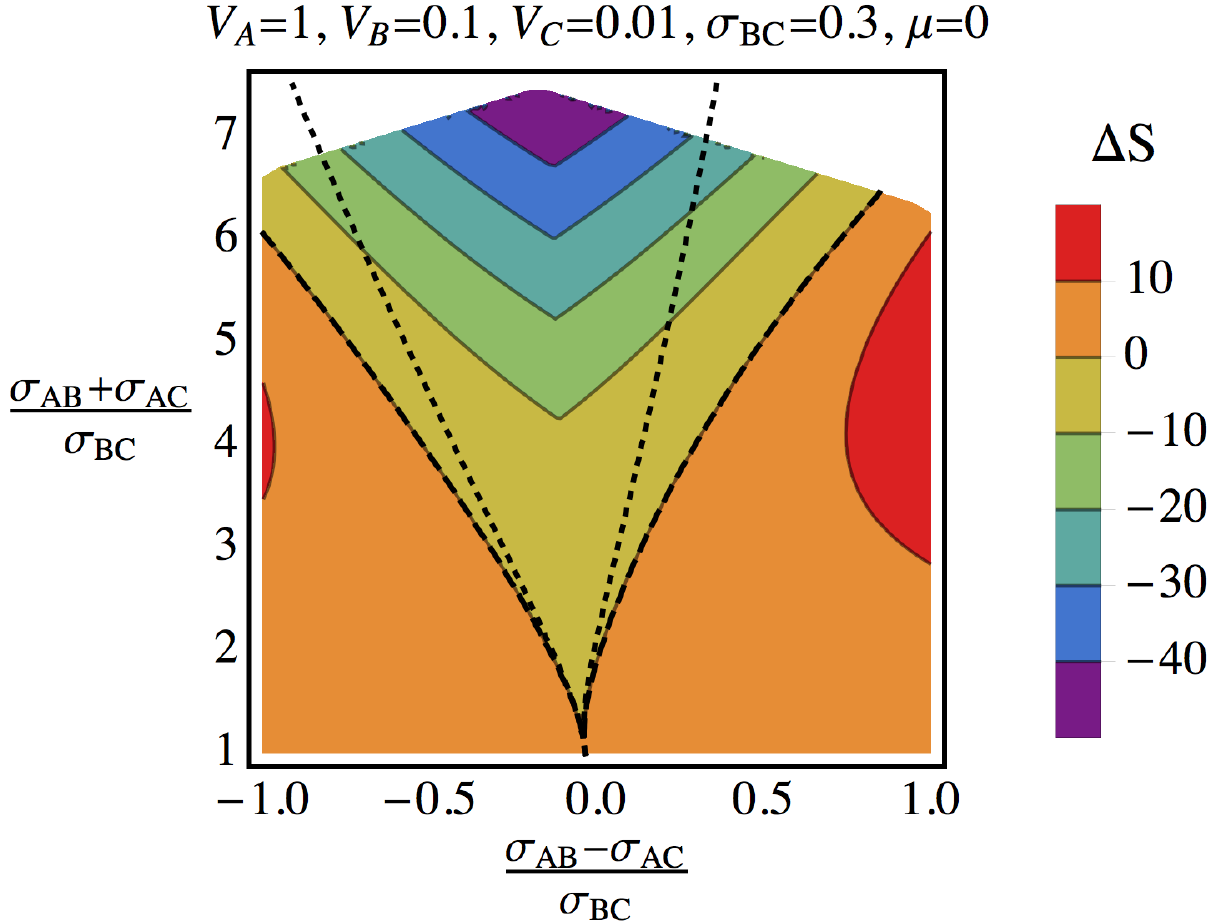}
\caption{$\Delta S = S_b - \mathrm{min}\left( 2S_{AB}, 2S_{AC}, S_{AB} + S_{AC}, S_{AB} + S_{BC} \right)$, \ie given an initial spherical bubble, the comparison between production of a barnacle or another spherical bubble, is plotted. The dashed line is the $\Delta S = 0$ contour, whilst the dotted is the same contour but neglecting the effects of gravity. The plot is bounded to the left, right and bottome due to the triangle inequalities \eqref{triangle inequality}, and at large $\sigma_{AB/C}$ it cuts off because $R_{AB/C}$ becomes too large.}\label{fig-DeltaS}
\end{figure}

First let us consider the effect of varying $\sigma_{AB}$ and $\sigma_{AC}$ whilst keeping everything else constant, \ie varying $S_{AB}$ and $S_{AC}$; in figure \ref{fig-DeltaS} is shown the difference of the barnacle action and the smallest of the set of actions for two spherical bubbles. One notes three things:
\begin{enumerate}[(i)]
\item as expected, when one of $S_{AB}$, $S_{AC}$ is much larger than the other (\ie right or left hand edges of the plot), the barnacle action is larger than an action involving two spherical bubbles;
\item in between these extremes, there is a region in which the barnacle action is less than any action involving two spherical bubbles, and as expected this region is centred around $S_{AB} \approx S_{AC}$;\footnote{Since $V_A - V_B$ is slightly less than $V_A - V_C$, this roughly corresponds to $\sigma_{AB}$ slightly less than $\sigma_{AC}$.}
\item the inclusion of gravity increases the region of parameter space in which barnacles are dominant (at least when gravity allows the existence barnacles).
\end{enumerate}
On the second point, it is worth emphasising that there is a region in which the barnacle action is smaller than the two spherical bubble actions by amounts $\mathcal{O}(10)$ and larger, and so even without calculating the exponential pre-factor one can be relatively confident that in this region the rate for barnacles is in fact dominant.

\begin{figure}[tp]
\centering
\includegraphics[scale=0.7]{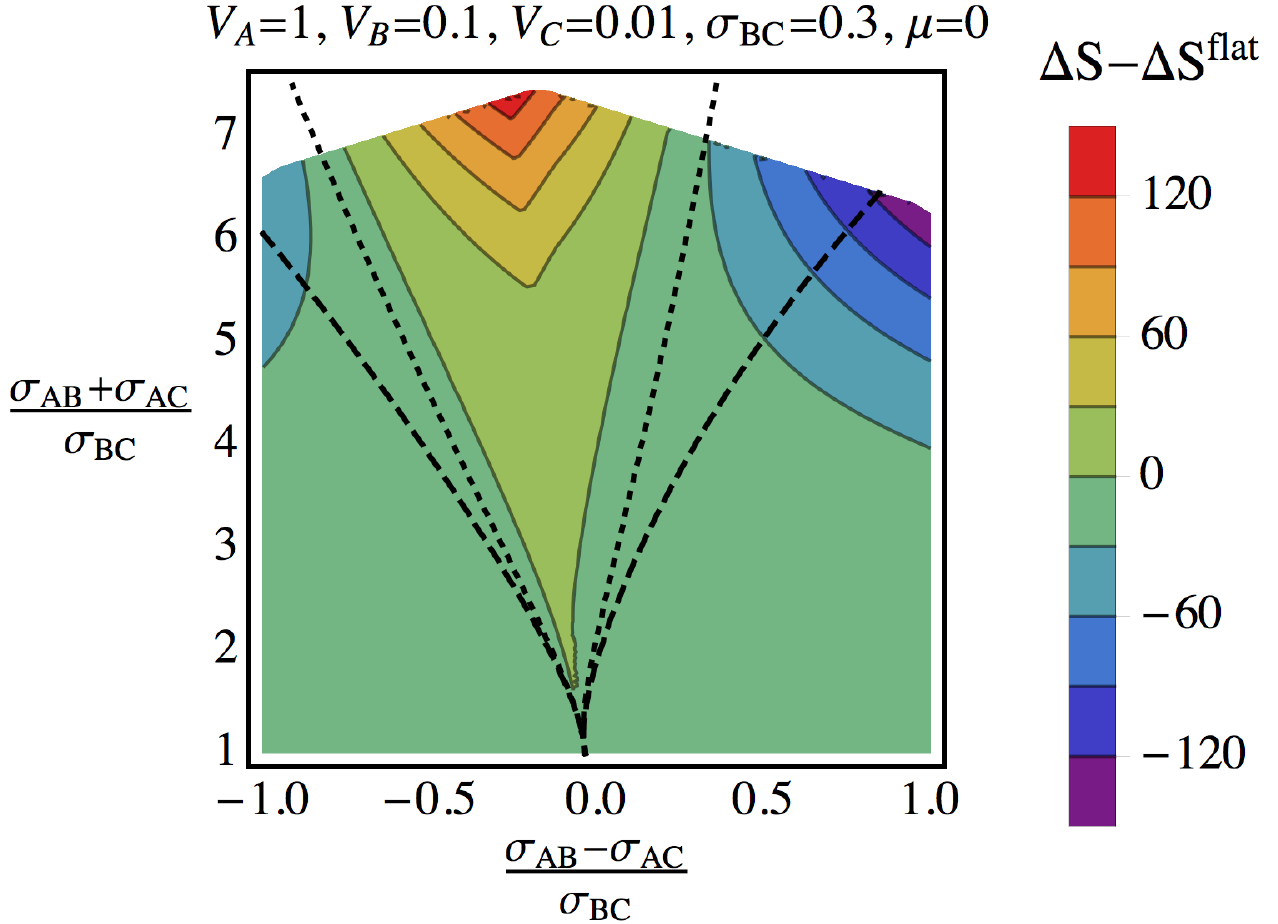}
\caption{$\Delta S = S_b - \mathrm{min}\left( 2S_{AB}, 2S_{AC}, S_{AB} + S_{AC}, S_{AB} + S_{BC} \right)$, and this figure plots the difference between this quantity with and without gravity. The dashed line is the $\Delta S = 0$ contour, whilst the dotted is $\Delta S^\text{flat} = 0$. The plot is bounded to the left, right and bottome due to the triangle inequalities \eqref{triangle inequality}, and at large $\sigma_{AB/C}$ it cuts off because $R_{AB/C}$ becomes too large.}\label{fig-DeltaDeltaS}
\end{figure}

To consider in more detail the third of the points mentioned above, examine figure \ref{fig-DeltaDeltaS}, which shows the difference of the quantity plotted in figure \ref{fig-DeltaS} and the same quantity for flat space. Interestingly one sees that gravity smooths things out: when barnacles are favoured, they are less favoured when gravity is included, and when barnacles are disfavoured, they are less disfavoured.

As mentioned in the introduction, often it is the case in classical nucleation that the heterogeneous rate dominates, since impurities can act as seeds for bubble formation. There have been studies of analogous effects in QFT, for example \cite{Grinstein:2015jda} studied in flat spacetime the effect of impurities much larger than than the bubble size, and found that they lead to a heterogeneous decay rate which can be either enhanced or suppressed relative to the standard homogeneous decay rate, depending on the details of the microphysics.

When gravity is included, there arises the possibility of black holes acting as nucleation sites; this has been studied by many authors, though most recently by \cite{Gregory:2013hja}. They find that the presence of a black hole below some critical mass (depending on the vacuum energy and putative bubble wall tension) leads to a decay rate which is enhanced with respect to the homogeneous case, which enhancement increases with black hole mass. Above the critical mass the enhancement decreases, until eventually the heterogeneous decay rate becomes smaller than the homogeneous one.

In the case of barnacles, one can think of the wall of an initial bubble acting as a seed for nucleation, and so it is interesting to compare with these previous results. Note that in figure \ref{fig-DeltaS} lines of constant $\sigma_{AB/C}$ are diagonal, and so as the wall tension of a spherical bubble increases, the favourability of barnacles increases---qualitatively similar to black hole mass. 

\begin{figure}[tp]
\centering
\includegraphics[scale=0.7]{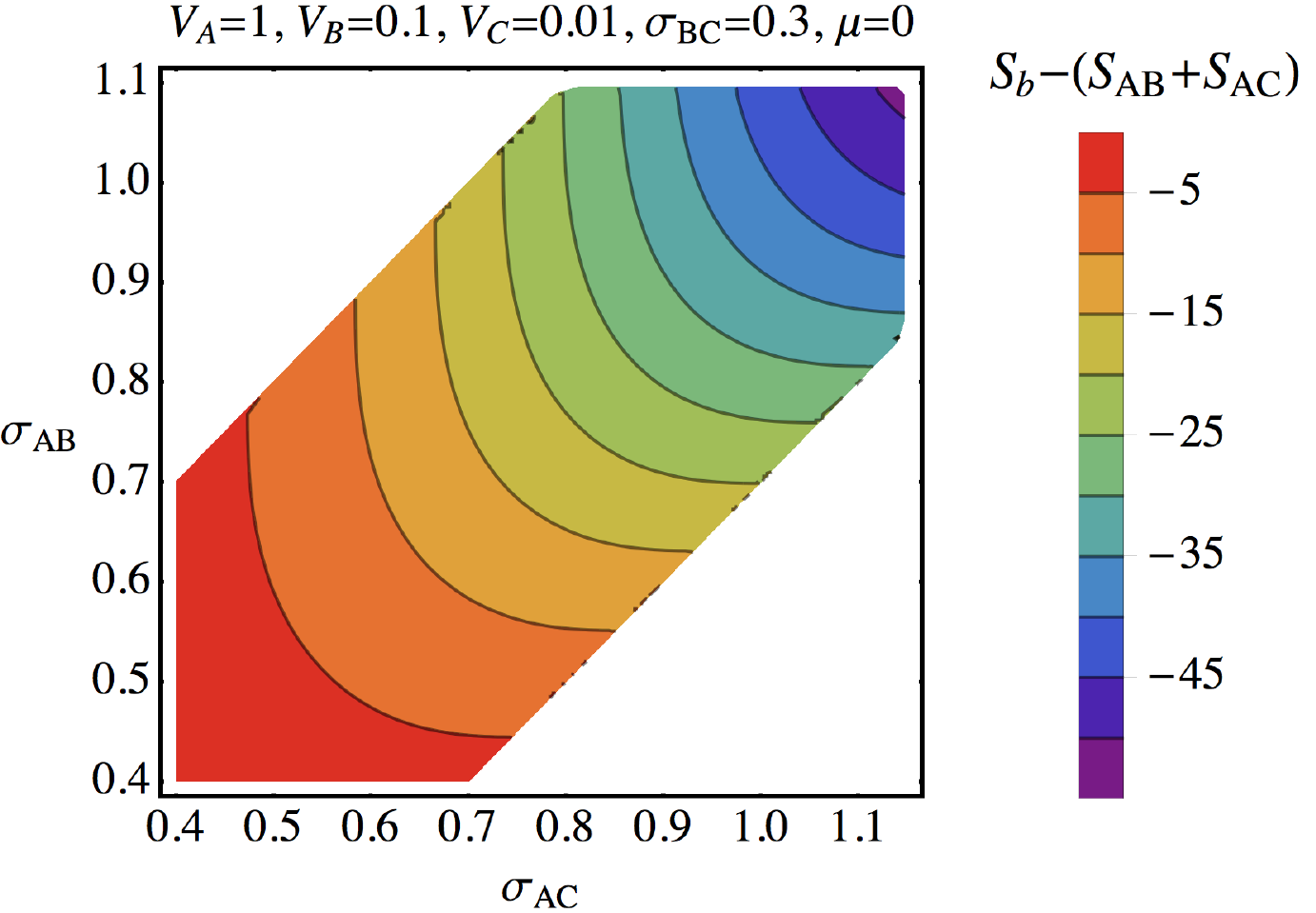}
\caption{Given an $AB/C$  bubble, this plot compares the rates for production of an $AC/B$ bubble to that of a barnacle, as a function of the $AB$ and $AC$ wall tensions. One sees that when the wall tensions are maximally different, the amount by which the barnacle is favoured is controlled mainly by the smaller wall tension (\ie the wall tension of the (probably) initial bubble).}\label{fig-Sb_minus_SAB_SAC}
\end{figure}

To make this more precise, let us compare $S_b$ to $S_{AB} + S_{AC}$ specifically, since then one is comparing the production of a spherical bubble of a certain type to the production of the barnacle produced by merging that bubble with the already existing bubble (whereas the action for two spherical bubbles of the same type does not have a direct barnacle limit). From the discussion of the previous subsection one already knows that the barnacle always has a lower action, however the precise behaviour is interesting; this is shown in figure \ref{fig-Sb_minus_SAB_SAC}. One sees that when the triangle inequalities are close to being saturated (\ie the edge of the plot) then the contours are orthogonal to increases in the smaller wall-tension. 

For example, consider $\sigma_{AB} \approx \sigma_{AC} + \sigma_{BC}$; this means $S_{AC} < S_{AB}$ and so it makes sense to consider the initial bubble to be $AC$; then figure \ref{fig-Sb_minus_SAB_SAC} tells us that the amount by which the wall of this bubble acts as a more efficient seed than empty $A$ vacuum for production of the $B$ vacuum depends mainly on the tension of the wall that is acting as the seed (rather than the tension of the wall that will be produced). In this sense, the wall tension is acting like the analogue of the black hole mass. Of course one key difference is that whereas for larger black hole masses the enhancement disappears, this is not the case for the wall tension (though one is of course constrained by the triangle inequalities). Similar behaviour is seen when comparing with $S_{AB} + S_{BC}$.

\subsubsection{$\mu$ Dependence}

\begin{figure}[tp]
\centering
\includegraphics[scale=0.7]{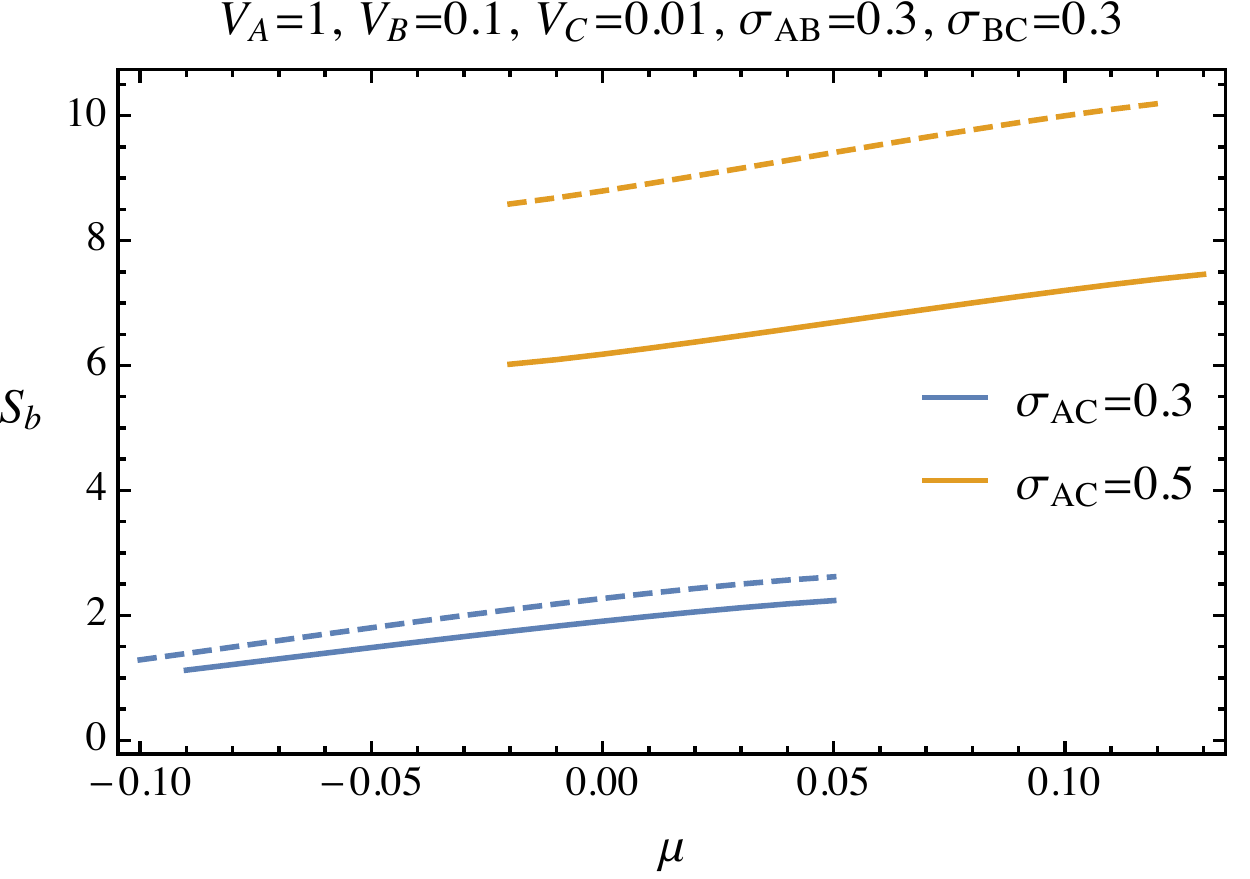}
\caption{The barnacle action as a function of $\mu$, the energy density at the junction point, with (solid) and without (dashed) gravity. Barnacles can only exist for $\mu$ which allow \eqref{conical singularity einstein} (with gravity) or \eqref{flat space r eqn} (without gravity) to be solved.}\label{fig-mu dependence}
\end{figure}

Thus far I have only considered $\mu = 0$ , and so in figure \ref{fig-mu dependence} one can see the dependence of the action on the energy density at the junction point, both with and without gravity, for two sets of parameters. As expected, positive $\mu$ increases the action, whilst negative $\mu$ decreases it; in particular, the barnacle action exhibits an approximately linear dependence on $\mu$. 

What is especially interesting are the similarities between the gravitational and flat-space cases, both in terms of which values of $\mu$ allow a barnacle to exist, and in terms of the behaviour of the action. This last is particularly curious since in the flat space case $\mu$ explicitly appears in the action once it has been extremised, but in the gravitational case it does not---see \eqref{Sbarn with grav - evaluate}.

\section{Phenomenological Applications and Observational Consequences} \label{sec-obs conqs}

\subsection{Bubble Collisions and Anisotropic Cosmologies}

In \cite{Czech:2011aa} it was noted that the nucleation of a barnacle on the wall of a bubble which contains an observer leads to identical observational consequences as the collision between the observer's bubble and another, since the consequences of the latter are just due to the $SO(1,2)$ symmetry of the collision hypersurface, which is precisely the symmetry of the barnacle---see \cite{Kleban:2011pg} for a brief review of cosmic bubble collisions.

Furthermore, assuming that the pre-factors in the decay rates are $\mathcal{O}(1)$, one can show that the number of barnacles an observer would expect to see in their past lightcone is larger than the number of collisions between bubbles of different vacua (and in some parameter regions is also larger than the number of collisions between bubbles of the same vacuum) \cite{Czech:2011aa}. Given the results of section \ref{sec-comp decay rates}, one sees that this conclusion holds when the effects of gravity are included.

Another intriguing prospect is that rather than living in a spherical bubble whose wall decays to barnacles, instead one could imagine living inside one of those barnacles. Just as the quantum state inside a bubble is modified from the usual Bunch-Davis vacuum on the scale of the bubble size, so too in the barnacle case one would expect modifications. In particular, the state would be anisotropic, and hence would lead to a background anisotropy in the power spectrum of primordial perturbations.\footnote{The effects of which can be studied in a similar manner to \cite{Scargill:2015bva}.} The scale on which this anisotropy is relevant can be smaller than the bubble size (and hence potentially observable) since the initial size of the barnacle in the direction along the axis of symmetry can be much smaller than the size perpendicular to it,\footnote{This is particularly the case when $\sigma_{AB} \approx \sigma_{AC} + \sigma_{BC}$ and $\sigma_{AC} \approx \sigma_{BC}$; following the discussion at the end of subsection \ref{sec-negative modes} such a barnacle would have to have a negative mode coming from variations of $\delta$ (the junction point position), whereas a barnacle with two negative modes coming from wall radii variations would lead to much milder anisotropy.} and furthermore it will in general not be parity invariant. One can also show that the lightcone of the origin of a barnacle must intersect one of the walls, inevitably leading to additional signatures. These points will be considered further in a future publication.

\subsection{Electroweak Baryogenesis}

Electroweak baryogenesis (EWBG) is the proposal to explain the observed baryon number asymmetry by using the electroweak phase transition in the early universe---see \cite{White:2016nbo} for a review. The essential ingredients for baryogenesis are given by the so-called Sakharov conditions: i) baryon number violation; ii) $C$ and $CP$ violation; iii) out-of-equilibrium dynamics; the latter of which can be effectively provided by a first order phase transition which proceeds via bubble nucleation. Unfortunately, in the standard model the electroweak phase transition is not first order, and so models of EWBG require additional ingredients.\footnote{It is also the case that the amount of $CP$ violation present in the standard model on its own (due to the phase in the CKM matrix) is not sufficient.}

Given the importance of bubble nucleation, one may wonder whether barnacles can play a role at all in EWBG---in particular, in a certain class of `two stage' models, such as studied in \cite{Patel:2012pi, Blinov:2015sna}, which involve a first order transition to an exotic electroweak symmetry breaking vacuum, in order to generate the baryon asymmetry, followed by a transition (which may or may not be first order) to the usual vacuum. 

In order for these models to be affected by barnacles it is clear that there must be a range of temperatures in which all the transitions between the three vacua (symmetric ($A$), exotic ($B$), and usual ($C$)) are first order.
In the models mentioned above one wants the $AB$ transition to percolate before $AC$ bubbles become significant, since by assumption $AC$ bubbles do not create enough baryon asymmetry (if they did, one could just use those and not need to consider a two stage model). To the extent that the results presented above may be valid at non-zero temperature, one sees that when $\Gamma_{AC}$ is not too much smaller than $\Gamma_{AB}$ it is possible that the walls of $AB$ bubbles decay to barnacles more quickly than new $AB$ bubbles are produced, which clearly impedes the completion of the $AB$ transition and thus baryogenesis. Hence barnacles may lead to additional constraints on some two stage models of EWBG.

\section{Conclusions} \label{sec-conc}

This paper has calculated the action for so-called `barnacles'---instantons which mediate the decay of a wall separating two regions of different vacuum energy---in the thin wall limit, whilst including the effects of gravity, thereby extending the previous calculation of \cite{Balasubramanian:2010kg}. 

Barnacles are described by seven parameters: three vacuum energies $V_i$, three wall tensions $\sigma_X$, and $\mu$, the energy density on the two-sphere at which the three vacua meet. In flat space, given $\{V_i, \sigma_X\}$, then a barnacle always exists when $\mu = 0$, but not necessarily if $|\mu|$ becomes too large. This largely remains the case when gravity is included, with the one caveat that the wall tensions must be small enough; if a wall tension does not satisfy the bound $V_f - V_t > \frac{3}{4}\kappa \sigma_{ft}^2$, then the radius of the segment of $ft$ bubble becomes too large, and precludes the existence of a stationary barnacle action. 

Another important observation is that whilst the inclusion of the $\mu$ parameter does not significantly affect the calculation of the flat space barnacle action, in the gravitational case it is imperative that it be considered, since it sources a deficit angle, and the contribution of this deficit angle to the action must be included if the correct solution is to be achieved upon making the action stationary.

Since a barnacle describes two decay processes: the production of the initial bubble, and the subsequent decay of its wall, it is appropriate to compare the barnacle to the action for the nucleation of two spherical bubbles. In this respect the inclusion of gravity does not change the qualitative picture much: the action for a barnacle action is smaller than all two bubble actions, except for those with two bubbles of the same type; the barnacle action is still smaller than the action for two bubbles of the same type, provided the rate for production of such spherical bubbles is not significantly greater than the rate for production of spherical bubbles of another type.
Gravity changes two things: it increases the the region of parameter space in which barnacles are favoured over same-type two bubble actions, and it mellows the behaviour so that both when a barnacle is favoured, and when it is disfavoured, the difference in actions (and hence rates) is less than in the flat space case.

Given the fact that barnacles can be competitive with spherical bubble nucleation, they may be phenomenologically relevant, for instance when considering creation of universes via tunnelling in a landscape (in which case gravitational effects may be especially important).
Previously, in \cite{Czech:2011aa}, it has been discussed how the observational signatures of barnacle may be similar to those of bubble collisions, and here it has also been noted that the anisotropy they induce in the primordial power spectrum, if one is inside a barnacle, may be relevant for cosmological observations, though this requires more investigation.

Other topics of further investigation include moving beyond the thin wall limit, and this would allow one to examine the question of how robust barnacles are to perturbations of the potential (both in terms of existence, and the value of their action). Another interesting extension would be to include the effects of finite temperature, especially as this may be relevant for electroweak baryogenesis. Finally, given progress in modelling random landscapes, for example see \cite{Bachlechner:2017zpb, Masoumi:2017gmh} for two recent approaches, it would be interesting to determine the likelihood and importance of barnacles in specific landscape models.
\\\\
\noindent{\bf Acknowledgements: }The author would like to thank Andreas Albrecht and Nemanja Kaloper for reading a draft of this paper; the author is supported by DOE grant DE-SC0009999.

\appendix
\section{Calculation of the Deficit Angle} \label{app-deficit angle calculation}

This appendix covers the details of the calculation of the deficit angle, \eqref{deficit angle}. Considering just $(\xi, \psi)$, the metric of the barnacle geometry can be written
\begin{subnumcases}{\mathrm{d}s^2 = }
\frac{3}{\kappa V_A} \left( \mathrm{d}\xi_A^2 + \sin^2 \xi_A \mathrm{d}\psi_A^2 \right) & \text{if } $\left\{
\begin{array}{cc}
0 \leq \psi_A < \frac{\pi}{2}, &\xi_{w,AC}(\psi_A) < \xi_A \leq \pi \\
\frac{\pi}{2} \leq \psi_A \leq \pi, &\xi_{w,AB}(\psi_A) < \xi_A \leq \pi
\end{array} \right. $ \label{A region} \\
\frac{3}{\kappa V_B} \left( \mathrm{d}\xi_B^2 + \sin^2 \xi_B \mathrm{d}\psi_B^2 \right) & \text{if } $\left\{
\begin{array}{c}
\frac{\pi}{2} \leq \psi_B \leq \pi, \\
\xi_{w,BC}(\psi_B) < \xi_B < \tilde{\xi}_{w,AB}(\psi_B) 
\end{array} \right. $ \label{B region} \\
\frac{3}{\kappa V_C} \left( \mathrm{d}\xi_C^2 + \sin^2 \xi_C \mathrm{d}\psi_C^2 \right) & \text{if } $\left\{
\begin{array}{cc}
0 \leq \psi_C < \frac{\pi}{2}, &0 \leq \xi_C < \tilde{\xi}_{w,AC}(\psi_C) \\
\frac{\pi}{2} \leq \psi_C \leq \pi, &0 \leq \xi_C < \tilde{\xi}_{w,BC}(\psi_C),
\end{array}
\right.$ \label{C region}
\end{subnumcases}
where $\xi_w$ gives the location of a bubble wall segment in the false vacuum coordinates, and $\tilde{\xi}_w$ the location in the true vacuum coordinates; explicit expressions for these, and the relationships between $\psi_{A,B,C}$ are derived below.

Consider embedding the $(\xi, \psi)$ coordinates in $\mathbb{R}^3$, and consider a circle on this two-sphere whose coordinate radius is $R$, and whose centre is displaced by a coordinate distance $\chi < R$ from the north pole. It is given by coordinates $(\xi_w(\psi), \psi)$, where
\begin{equation}
\begin{pmatrix}
\sin \xi_w \cos \psi \\
\sin \xi_w \sin \psi \\
\cos \xi_w
\end{pmatrix}
=
\begin{pmatrix}
\cos \chi & 0 & -\sin \chi \\
0 & 1 & 0 \\
\sin \chi & 0 & \cos \chi
\end{pmatrix}
\begin{pmatrix}
\sin R \cos \tilde{\psi} \\
\sin R \sin \tilde{\psi} \\
\cos R
\end{pmatrix}
=
\begin{pmatrix}
\cos \chi \sin R \cos \tilde{\psi} - \sin \chi \cos R \\
\sin R \sin \tilde{\psi} \\
\sin \chi \sin R \cos \tilde{\psi} + \cos \chi \cos R
\end{pmatrix},
\end{equation}
which can be solved to give
\begin{align}
\sin \xi_w &= \frac{\cos R \sin \chi \cos \psi + \cos \chi \sqrt{\sin^2 R - \sin^2 \chi \sin^2 \psi}}{1 - \sin^2 \chi \sin^2 \psi} \label{sin xi w} \\
\cos \xi_w &= \frac{\cos R \cos \chi - \sin \chi \cos \psi \sqrt{\sin^2 R - \sin^2 \chi \sin^2 \psi}}{1 - \sin^2 \chi \sin^2 \psi}. \label{cos xi w}
\end{align}
It can be verified that these give the expected answers $\xi_w(0) = R + \chi$ and $\xi_w(\pi) = R - \chi$, and (numerically) that they give the correct answer for the volume of the bubble segment, \ie 
\begin{equation}
4\pi \int_0^{\frac{\pi}{2}} \mathrm{d}\psi\, \sin^2 \psi \int_0^{\xi_w} \mathrm{d}\xi\, \sin^3 \xi = 4\pi \int_0^{\frac{\pi}{2}} \mathrm{d}\psi\, \sin^2 \psi \left( \frac{2}{3} - \cos\xi_w + \frac{1}{3}\cos^3\xi_w \right),
\end{equation}
equals \eqref{bubble segment volume}.

Due to the fact that the true and false vacuum four spheres have different radii, the planes in the embedding space which go through the junction point and contain the centres of the spheres are misaligned by the angle $\theta$ calculated in section \ref{sec-theta}. This means that the azimuthal angular coordinates of the true and false vacuum coordinate systems are not equal.

Again, consider embedding $(\xi, \psi)$ in three dimensional Euclidean space, and align one axis along $(0,0,1)$, and the other along $(\sin \theta, 0, \cos \theta)$; in doing so, the junction has been placed at $z = 0$, and so the bubble wall has coordinates 
\begin{equation}
\begin{pmatrix}
\sin \xi_w \cos \psi \\
\sin \xi_w \sin \psi \\
\underbrace{\cos \xi_w - \cos \xi_w\left(\psi=\frac{\pi}{2}\right)}_{-\sin \xi_w \cos \psi \tan \chi}
\end{pmatrix}.
\end{equation}
The perpendicular from the $(\sin \theta, 0, \cos \theta)$ axis to the bubble wall is
\begin{equation}
\begin{pmatrix}
\sin \xi_w \cos \psi \left[ 1 + \sin \theta \left( \cos \theta \tan \chi - \sin \theta \right) \right] \\
\sin \xi_w \sin \psi \\
\sin \xi_w \cos \psi \left[ - \tan \chi + \cos \theta \left( \cos \theta \tan \chi - \sin \theta \right) \right]
\end{pmatrix},
\end{equation}
and the length of this gives the matching condition to ensure the metric is continuous at the bubble wall:
\begin{equation}
\frac{3}{\kappa V_B} \sin^2 \xi_B = \frac{3}{\kappa V_A} \sin^2 \xi_w \left( \sin^2 \psi + \cos^2 \psi \frac{\cos^2 (\chi - \theta)}{\cos^2 \chi} \right). \label{xi tilde xi relation}
\end{equation}
On the right hand side the azimuthal angle around the $(0,0,1)$ axis, $\psi$, should be understood to be function of $\tilde{\psi}$, the azimuthal angle around the $(\sin \theta, 0, \cos \theta)$ axis. Taking this to be measured starting from the direction $(\cos \theta, 0, -\sin \theta)$ one has
\begin{equation}
\cos \tilde{\psi} = \frac{\cos \psi \cos(\chi - \theta)}{\sqrt{\sin^2 \psi \cos^2 \chi + \cos^2 \psi \cos^2 (\chi - \theta)}}. \label{psi tilde psi relation}
\end{equation}

With these expressions in hand, we can now go on to calculate the deficit angle. The junction point is located at
\begin{equation}
\sin \xi_i = \sqrt{\frac{V_i}{V_A}} \sin \delta, \qquad \psi_i = \frac{\pi}{2},
\end{equation}
and the deficit angle around it can be calculated by transforming to a coordinate system which at this location locally looks like flat space in polar coordinates, \ie $\mathrm{d}s^2 = \mathrm{d}r^2 + r^2 \mathrm{d}\phi^2$, and then determining the range of the angular variable $\phi$. For the $A$ region this transformation takes the form
\begin{equation}
\xi_A = \delta + \sqrt{\frac{\kappa V_A}{3}} r \cos \phi, \qquad \psi_A = \frac{\pi}{2} + \sqrt{\frac{\kappa V_A}{3}} \frac{1}{\sin \delta} r \sin \phi,
\end{equation}
which gives
\begin{equation}
\sin \xi_A = \sin \delta + \cos \delta \sqrt{\frac{\kappa V_A}{3}} r \cos \phi + \mathcal{O}(r^2),
\end{equation}
and from \eqref{sin xi w} one has
\begin{equation}
\sin \xi_{w,Aj} = \sin \delta \mp \tan \chi_{Aj} \frac{\cos R_{Aj}}{\cos \chi_{Aj}} \sqrt{\frac{\kappa V_A}{3}} \frac{1}{\sin \delta} r \sin \phi + \mathcal{O}(r^2),
\end{equation}
where the upper sign is for $AC$ and the lower for $AB$.
Finally, noting that $\frac{\cos R_{Aj}}{\cos \chi_{Aj}} = \cos \delta$, in the vicinity of the junction point the range of \eqref{A region} becomes 
\begin{equation}
- \cot^{-1} \left( \frac{\tan \chi_{AC}}{\sin \delta} \right) < \phi < \cot^{-1} \left( \frac{\tan \chi_{AB}}{\sin \delta} \right), \label{A range}
\end{equation}
where the range of the inverse cotangent is taken to be $(0,\pi)$.

For the $B$ region one has
\begin{equation}
\xi_B = \sin^{-1} \left( \sqrt{\frac{V_B}{V_A}} \sin \delta \right) + \sqrt{\frac{\kappa V_B}{3}} r \cos \phi, \qquad \psi_B = \frac{\pi}{2} + \sqrt{\frac{\kappa V_B}{3}} \frac{1}{ \sqrt{\frac{V_B}{V_A}} \sin \delta} r \sin \phi,
\end{equation}
and hence
\begin{equation}
\sin \xi_B = \sqrt{\frac{V_B}{V_A}} \sin \delta - \sqrt{1 - \frac{V_B}{V_A}\sin^2 \delta } \sqrt{\frac{\kappa V_B}{3}} r \cos \phi + \mathcal{O}(r^2).
\end{equation}
The $BC$ boundary can be dealt with in much the same way as the boundaries of the $A$ region, and one has
\begin{equation}
\sin \xi_{w,Aj} = \sqrt{\frac{V_B}{V_A}} \sin \delta + \tan \chi_{BC} \frac{\cos R_{BC}}{\cos \chi_{BC}} \sqrt{\frac{\kappa V_B}{3}} \frac{1}{ \sqrt{\frac{V_B}{V_A}} \sin \delta} r \sin \phi + \mathcal{O}(r^2),
\end{equation}
and note that $\frac{\cos R_{BC}}{\cos \chi_{BC}} = \sqrt{1 - \frac{V_B}{V_C} \sin^2 \delta}$. The $AB$ boundary requires one to relate $\psi_B$ to $\psi_A$, which is done in \eqref{psi tilde psi relation}, and to formulate the correct matching condition at the wall, which is done in \eqref{xi tilde xi relation}, with the result
\begin{equation}
\sin \tilde{\xi}_{w,AB} = \sqrt{\frac{V_B}{V_A}} \left( \sin \delta + \tan \chi_{AB} \frac{\cos R_{AB}}{\cos \chi_{AB}} \frac{\cos \chi_{AB}}{\cos(\chi_{AB} - \theta_{AB})} \sqrt{\frac{\kappa V_B}{3}} \frac{1}{ \sqrt{\frac{V_B}{V_A}} \sin \delta} r \sin \phi + \mathcal{O}(r^2) \right).
\end{equation}
After a few algebraic manipulations, the range of \eqref{B region} in the vicinity of the junction point then becomes 
\begin{equation}
\cot^{-1} \left(\frac{\tan (\chi_{AB} - \theta_{AB})}{ \sqrt{\frac{V_B}{V_A}} \sin \delta} \right) < \phi < \cot^{-1} \left( \frac{\tan \chi_{BC}}{\sqrt{\frac{V_B}{V_A}} \sin \delta} \right). \label{B range}
\end{equation}

The boundaries of the $C$ region can be dealt with in much the same way as the $AB$ boundary of the $B$ region, with the result that in the vicinity of the junction point the range of \eqref{C region} becomes 
\begin{equation}
2\pi - \cot^{-1} \left(\frac{\tan (\chi_{AC} - \theta_{AC})}{ \sqrt{\frac{V_C}{V_A}} \sin \delta} \right) < \phi < \cot^{-1} \left( \frac{\tan (\chi_{BC} - \theta_{BC})}{\sqrt{\frac{V_C}{V_A}} \sin \delta} \right). \label{C range}
\end{equation}

Finally, the deficit angle is given by $2\pi$ minus the sum of the angular extents \eqref{A range}, \eqref{B range}, and \eqref{C range}:
\begin{align}
\Delta = &\cot^{-1} \left(\frac{\tan (\chi_{AB} - \theta_{AB})}{ \sqrt{\frac{V_B}{V_A}} \sin \delta} \right) - \cot^{-1} \left( \frac{\tan \chi_{AB}}{\sin \delta} \right) \nonumber \\
&+ \cot^{-1} \left(\frac{\tan (\chi_{AC} - \theta_{AC})}{ \sqrt{\frac{V_C}{V_A}} \sin \delta} \right) - \cot^{-1} \left( \frac{\tan \chi_{AC}}{\sin \delta} \right) \nonumber \\
&+ \cot^{-1} \left( \frac{\tan (\chi_{BC} - \theta_{BC})}{\sqrt{\frac{V_C}{V_A}} \sin \delta} \right) - \cot^{-1} \left( \frac{\tan \chi_{BC}}{\sqrt{\frac{V_B}{V_A}} \sin \delta} \right).
\end{align}

\bibliographystyle{JHEP}
\bibliography{Barnacles_and_Gravity}

\providecommand{\href}[2]{#2}\begingroup\raggedright\begin{thebibliography}{10}

\bibitem{Coleman:1977py}
S.~R. Coleman, \emph{{The Fate of the False Vacuum. 1. Semiclassical Theory}},
  \href{http://dx.doi.org/10.1103/PhysRevD.15.2929}{\emph{Phys.Rev.} {\bf D15}
  (1977) 2929--2936}.

\bibitem{Callan:1977pt}
J.~Callan, Curtis~G. and S.~R. Coleman, \emph{{The Fate of the False Vacuum. 2.
  First Quantum Corrections}},
  \href{http://dx.doi.org/10.1103/PhysRevD.16.1762}{\emph{Phys.Rev.} {\bf D16}
  (1977) 1762--1768}.

\bibitem{Coleman:1980aw}
S.~R. Coleman and F.~De~Luccia, \emph{{Gravitational Effects on and of Vacuum
  Decay}}, \href{http://dx.doi.org/10.1103/PhysRevD.21.3305}{\emph{Phys.Rev.}
  {\bf D21} (1980) 3305}.

\bibitem{Hawking:1981fz}
S.~W. Hawking and I.~G. Moss, \emph{{Supercooled Phase Transitions in the Very
  Early Universe}},
  \href{http://dx.doi.org/10.1016/0370-2693(82)90946-7}{\emph{Phys. Lett.} {\bf
  B110} (1982) 35--38}.

\bibitem{Brown:2007sd}
A.~R. Brown and E.~J. Weinberg, \emph{{Thermal derivation of the Coleman-De
  Luccia tunneling prescription}},
  \href{http://dx.doi.org/10.1103/PhysRevD.76.064003}{\emph{Phys. Rev.} {\bf
  D76} (2007) 064003}, [\href{http://arxiv.org/abs/0706.1573}{{\tt
  0706.1573}}].

\bibitem{Coleman:1977th}
S.~R. Coleman, V.~Glaser and A.~Martin, \emph{{Action Minima Among Solutions to
  a Class of Euclidean Scalar Field Equations}},
  \href{http://dx.doi.org/10.1007/BF01609421}{\emph{Commun. Math. Phys.} {\bf
  58} (1978) 211}.

\bibitem{Blum:2016ipp}
K.~Blum, M.~Honda, R.~Sato, M.~Takimoto and K.~Tobioka, \emph{{O($N$)
  Invariance of the Multi-Field Bounce}},
  \href{http://arxiv.org/abs/1611.04570}{{\tt 1611.04570}}.

\bibitem{Grinstein:2015jda}
B.~Grinstein and C.~W. Murphy, \emph{{Semiclassical Approach to Heterogeneous
  Vacuum Decay}}, \href{http://dx.doi.org/10.1007/JHEP12(2015)063}{\emph{JHEP}
  {\bf 12} (2015) 063}, [\href{http://arxiv.org/abs/1509.05405}{{\tt
  1509.05405}}].

\bibitem{Gregory:2013hja}
R.~Gregory, I.~G. Moss and B.~Withers, \emph{{Black holes as bubble nucleation
  sites}}, \href{http://dx.doi.org/10.1007/JHEP03(2014)081}{\emph{JHEP} {\bf
  03} (2014) 081}, [\href{http://arxiv.org/abs/1401.0017}{{\tt 1401.0017}}].

\bibitem{Balasubramanian:2010kg}
V.~Balasubramanian, B.~Czech, K.~Larjo and T.~S. Levi, \emph{{Vacuum decay in
  multidimensional field landscapes: thin, thick and intersecting walls}},
  \href{http://dx.doi.org/10.1103/PhysRevD.86.049904,
  10.1103/PhysRevD.84.025019}{\emph{Phys. Rev.} {\bf D84} (2011) 025019},
  [\href{http://arxiv.org/abs/1012.2065}{{\tt 1012.2065}}].

\bibitem{Czech:2011aa}
B.~Czech, \emph{{A Novel Channel for Vacuum Decay}},
  \href{http://dx.doi.org/10.1016/j.physletb.2012.06.018}{\emph{Phys. Lett.}
  {\bf B713} (2012) 331--334}, [\href{http://arxiv.org/abs/1112.1638}{{\tt
  1112.1638}}].

\bibitem{Coleman:1987rm}
S.~R. Coleman, \emph{{Quantum Tunneling and Negative Eigenvalues}},
  \href{http://dx.doi.org/10.1016/0550-3213(88)90308-2}{\emph{Nucl. Phys.} {\bf
  B298} (1988) 178--186}.

\bibitem{Fursaev:1995ef}
D.~V. Fursaev and S.~N. Solodukhin, \emph{{On the description of the Riemannian
  geometry in the presence of conical defects}},
  \href{http://dx.doi.org/10.1103/PhysRevD.52.2133}{\emph{Phys. Rev.} {\bf D52}
  (1995) 2133--2143}, [\href{http://arxiv.org/abs/hep-th/9501127}{{\tt
  hep-th/9501127}}].

\bibitem{Hayward:1993my}
G.~Hayward, \emph{{Gravitational action for space-times with nonsmooth
  boundaries}}, \href{http://dx.doi.org/10.1103/PhysRevD.47.3275}{\emph{Phys.
  Rev.} {\bf D47} (1993) 3275--3280}.

\bibitem{Israel:1966rt}
W.~Israel, \emph{{Singular hypersurfaces and thin shells in general
  relativity}}, \href{http://dx.doi.org/10.1007/BF02710419,
  10.1007/BF02712210}{\emph{Nuovo Cim.} {\bf B44S10} (1966) 1}.

\bibitem{Vilenkin:1981zs}
A.~Vilenkin, \emph{{Gravitational Field of Vacuum Domain Walls and Strings}},
  \href{http://dx.doi.org/10.1103/PhysRevD.23.852}{\emph{Phys. Rev.} {\bf D23}
  (1981) 852--857}.

\bibitem{Parke:1982pm}
S.~J. Parke, \emph{{Gravity, the Decay of the False Vacuum and the New
  Inflationary Universe Scenario}},
  \href{http://dx.doi.org/10.1016/0370-2693(83)91376-X}{\emph{Phys. Lett.} {\bf
  121B} (1983) 313--315}.

\bibitem{Kleban:2011pg}
M.~Kleban, \emph{{Cosmic Bubble Collisions}},
  \href{http://dx.doi.org/10.1088/0264-9381/28/20/204008}{\emph{Class. Quant.
  Grav.} {\bf 28} (2011) 204008}, [\href{http://arxiv.org/abs/1107.2593}{{\tt
  1107.2593}}].

\bibitem{Scargill:2015bva}
J.~H.~C. Scargill, \emph{{An anisotropic universe due to dimension-changing
  vacuum decay}},
  \href{http://dx.doi.org/10.1088/1475-7516/2015/08/045}{\emph{JCAP} {\bf 1508}
  (2015) 045}, [\href{http://arxiv.org/abs/1506.07100}{{\tt 1506.07100}}].

\bibitem{White:2016nbo}
G.~A. White, \emph{{A Pedagogical Introduction to Electroweak Baryogenesis}}.
\newblock IOP Concise Physics. Morgan and Claypool, 2016,
  \href{http://dx.doi.org/10.1088/978-1-6817-4457-5}{10.1088/978-1-6817-4457-5}.

\bibitem{Patel:2012pi}
H.~H. Patel and M.~J. Ramsey-Musolf, \emph{{Stepping Into Electroweak Symmetry
  Breaking: Phase Transitions and Higgs Phenomenology}},
  \href{http://dx.doi.org/10.1103/PhysRevD.88.035013}{\emph{Phys. Rev.} {\bf
  D88} (2013) 035013}, [\href{http://arxiv.org/abs/1212.5652}{{\tt
  1212.5652}}].

\bibitem{Blinov:2015sna}
N.~Blinov, J.~Kozaczuk, D.~E. Morrissey and C.~Tamarit, \emph{{Electroweak
  Baryogenesis from Exotic Electroweak Symmetry Breaking}},
  \href{http://dx.doi.org/10.1103/PhysRevD.92.035012}{\emph{Phys. Rev.} {\bf
  D92} (2015) 035012}, [\href{http://arxiv.org/abs/1504.05195}{{\tt
  1504.05195}}].

\bibitem{Bachlechner:2017zpb}
T.~C. Bachlechner, K.~Eckerle, O.~Janssen and M.~Kleban, \emph{{Axions of
  Evil}},  \href{http://arxiv.org/abs/1703.00453}{{\tt 1703.00453}}.

\bibitem{Masoumi:2017gmh}
A.~Masoumi, A.~Vilenkin and M.~Yamada, \emph{{Initial conditions for slow-roll
  inflation in a random Gaussian landscape}},
  \href{http://arxiv.org/abs/1704.06994}{{\tt 1704.06994}}.

\end{thebibliography}\endgroup

\end{document}